\documentclass[twocolumn,twocolappendix]{aastex631}

\usepackage{amsmath,comment}

\newcommand{\ppdot}{\mbox{$P$-$\dot P$}}
\newcommand{\pdot}{\mbox{$\dot P$}}
\newcommand{\edot}{\mbox{$\dot E$}}
\newcommand{\ergs}{\,erg\,s$^{-1}$}
\newcommand{\kms}{\,km\,s$^{-1}$}

\received{2024 March 25}
\revised{2024 May 21}
\accepted{2024 May 27}
\submitjournal{ApJ}
\begin{document}
\title{Pulsar Population Synthesis with Magneto-Rotational Evolution: Constraining the Decay of Magnetic Field}
\author[0000-0002-1904-4957]{Zhihong Shi}
\affiliation{Department of Physics, The University of Hong Kong, Pokfulam Rd., Hong Kong, China}
\author[0000-0002-5847-2612]{C.-Y. Ng}
\affiliation{Department of Physics, The University of Hong Kong, Pokfulam Rd., Hong Kong, China}


\shorttitle{Pulsar population with magneto-rotational evolution}

\begin{abstract}
We present a population synthesis model for normal radio pulsars in the Galaxy, incorporating the latest developments in the field and the magneto-rotational evolution processes. Our model considers spin-down with force-free magnetosphere and the decay of the magnetic field strength and its inclination angle. The simulated pulsar population is fit to a large observation sample that covers the majority of radio surveys using Markov Chain Monte Carlo (MCMC) technique. We compare the distributions of four major observables: spin period ($P$), spin-down rate, (\pdot), dispersion measure (DM) and radio flux density, using accurate high-dimensional Kolmogorov-Smirnov (KS) statistic. We test two $B$-field decay scenarios, an exponential model motivated by Ohmic dissipation and a power-law model motivated by the Hall effect. The former clearly provides a better fit and it can successfully reproduce the observed pulsar distributions with a decay time scale of $8.3^{+3.9}_{-3.0}$\,Myr. The result suggests significant $B$-field decay in aged pulsars and Ohmic dissipation could be the dominant process.
\end{abstract}

\section{Introduction}
Since the first pulsar was detected in 1967 \citep{hewish1969observation}, to date there are over 3000 known pulsars in the sky according to the Australia Telescope National Facility (ATNF) catalog \citep{manchester2005australia}. They are very interesting objects and powerful tools for probing many different aspects of physics, from relativity to gravity. 
For example, understanding their initial spin period ($P_0$) and magnetic field ($B_0$) distributions could reveal the angular momentum transport process and the origin of $B$-field during core collapse. Modeling the pulsar population could provide tests to the emission mechanisms and constrain the event rates for the related high-energy astrophysical phenomena, including supernovae, fast radio bursts, gravitational waves, and Gamma-ray bursts. 

Normal pulsars lose their rotational energy in the form of particle wind and electromagnetic radiation, which results in spin-down. The spin parameters ($P$ and \pdot) thus depend on the $B$-field strength and configuration. The evolution of the magnetic field could leave observable signatures on the pulsar rotational properties. This is the so-called
magneto-rotational evolution scenario \citep[see][]{gullon2014population}.

That being said, the detailed evolution of pulsar $B$-field remains an open question. There is indirect evidence that suggests magnetic field decay during the pulsar lifetime. For instance, the strongest $B$-fields are found in young pulsars and weak fields are found in old millisecond pulsars (MSPs), which is generally consistent with the idea of field decay expected from theories \citep[e.g.][]{goldreich1992magnetic,igoshev2021evolution}.
Moreover, some pulsars are found to have braking index deviated from 3, which could be the result of decreasing magnetic field strength \citep[e.g.,][]{vigano2013unifying, igoshev2019ages} due to Ohmic or Hall effects \citep{flowers1977evolution,goldreich1992magnetic,igoshev2015magnetic,igoshev2019ages} or evolution of the magnetic alignment angle \citep[e.g.,][]{weltevrede2008population} by plasma effect \citep{philippov2014time}.

Different theories suggest a wide range of decay time scales from a few $10^5$\,yr \citep{igoshev2015magnetic} to 150\,Myr \citep{bransgrove2018magnetic}. Observationally, while the decay rate is difficult to measure directly, there are previous attempts to obtain constraints from population modeling. Most of the reported values are in the range of a few Myr \citep[e.g.][]{gonthier2004role,cieslar2020markov}, but some found no
evidence of decay \citep[e.g.,][]{faucher2006birth}. In this study, we revisit this problem by developing a state-of-the-art population model that includes the latest developments in the field as realistic physics inputs, and fit it to a large pulsar sample to constrain the model parameters.
Our model adopts the spin-down formula in a plasma-filled magnetosphere \citep{spitkovsky2006time,philippov2014time} with the explicit dependence of $\alpha$, the updated kick velocity distribution \citep{verbunt2017observed,igoshev2020observed}, sophisticated radio beam model \citep{gonthier2018population}, an updated free electron density distribution of the Galaxy 
\citep{yao2017new}, and  the latest Gamma-ray luminosity law and beam geometry \citep{watters2009atlas,kalapotharakos2017fermi,kalapotharakos2019fundamental,kalapotharakos2022fundamental}.
In previous works, model comparison is often only limited to pulsars from the Parkes and Swinburne Multibeam (PMB) surveys \citep{manchester2001parkes,edwards2001swinburne} and the High Time Resolution Universe (HTRU) survey \citep{keith2010high}.
We attempt to fit to a much larger observation sample that covers all major normal radio surveys.
Finally, we employ, for the first time, a correct formula for the KS statistic at high dimension for model comparison
\citep{fasano1987multid,justel1997multivariate}.

This paper is organized as follows: in Section~\ref{s2}, we introduce the simulation methods including five parts: birth properties, evolution, emission, detection, and comparison with observations. In Section~\ref{sec:result}, we report our simulation results for the optimal parameters and the corresponding population properties. The discussion is in Section~\ref{s4} and the conclusion is drawn in Section~\ref{s5}.

\section{Simulation method\label{s2}}
\subsection{Overview of the simulation and fitting procedure}
We adopt the evolution approach for the population synthesis. We follow the basic procedures outlined in \citet{faucher2006birth} but with more realistic physics and model
inputs from recent studies.
First, we assign birth properties to newborn pulsars, i.e.\ $P_0$, $B_0$, $\alpha$, location, kick velocity and direction, and viewing angle.
Then each pulsar starts two independent evolution processes: magneto-rotational evolution and dynamical motion under the Galactic gravitational potential.
The evolution time of each pulsar is randomly assigned from a uniform distribution between 0 and 100\,Myr.
Since it is computationally more intensive to trace the pulsar motion in the Galaxy,
we first calculate the evolution of $P$, \pdot, and $\alpha$, and use the final values 
to determine if a pulsar is visible. We assume a distance of 1\,pc and set a
threshold of $10^{-6}$\,mJy\,MHz, which is much lower than the sensitivity of
any radio surveys used in this study.
If a pulsar is beamed away from the Earth or has radio flux below the above limit, we can safely assume that it cannot be detected and move on to simulate another pulsar.
Only for pulsars pass this initial screening we calculate their orbits.
We stress that this procedure has no effect on the simulated pulsar population
since the two evolutions are independent.
At the end of the simulation, we model the radio and Gamma-ray emission based on their final spin property, location, and viewing geometry.
A pulsar is only counted if it is detectable by the radio surveys in our list.


In each simulation run, we generate 10 times more detectable pulsars than the
observation sample (i.e.\ a total of 18590 pulsars). This number was chosen to
minimize the statistical fluctuation while keeping the simulation efficient. See
Appendix~\ref{Appendix:B} for more details. After each run, we then determine
the goodness of fit for a given set of model parameters, by comparing the
simulated pulsar distribution with the observations using a high-dimensional
Kolmogorov-Smirnov (KS) statistic. Finally, based on the KS statistic, we fit
the model parameters using the Markov chain Monte Carlo (MCMC) technique.

\subsection{Birth properties}
\subsubsection{Spatial distribution}
Normal pulsars are believed to be born in core-collapse supernova of massive stars.
Due to short lifetime of the massive stars, the progenitors are not expected to travel far before their birthplace.
We can therefore assume that pulsars are born very close to the spiral arms of the Galaxy.
We adopt a logarithmic function to depict the four spiral arms of the Milky Way
\begin{equation}
    \theta(r_{xy})=k\ln{(r_{xy}/r_0)} +\theta_0, \label{eqn:arm}
\end{equation}
where $k$ and $\theta_0$ are constants from \citet{wainscoat1992model}, 
$r_{xy}=\sqrt{x^2+y^2}$, $\theta=\arctan(y/x)$, where $x,y,z$ are coordinate values in a left-handed Galactocentric Cartesian coordinate system such that the Sun locates at $(x,y,z)=(8.5,0,0.025)$\,kpc. 
We then adopt the radial distribution model of surface density ($\rho(r_{xy})$) for newborn pulsars
\begin{equation}
    \rho(r_{xy})\propto r_{xy} \left(\frac{r_{xy}+R_1}{R_\odot+R_1}\right)^a\exp\left[-b\left(
    \frac{r_{xy}-R_\odot}{R_\odot+R_1}\right)\right],
\end{equation}
where $R_\odot$ is the distance from the Sun to the Galactic center, and $a$, $b$,
and $R_1$ are constants from \citet{yusifov2004revisiting}.
Next, we apply a polar angle correction $\theta_{\rm corr}$ and distance adjustment $r_{\rm corr}$ to each pulsar following \cite{faucher2006birth} to blur the pulsar distribution,
so that they are not all piled up at the centroids of arms.
Finally, a two-sided exponential function is applied to describe the distribution of vertical distance above and below the Galactic plane with a scale height $z_0$ of 50\,pc

\begin{figure}[tbh]
    \centering
    \includegraphics[width=0.49\textwidth]{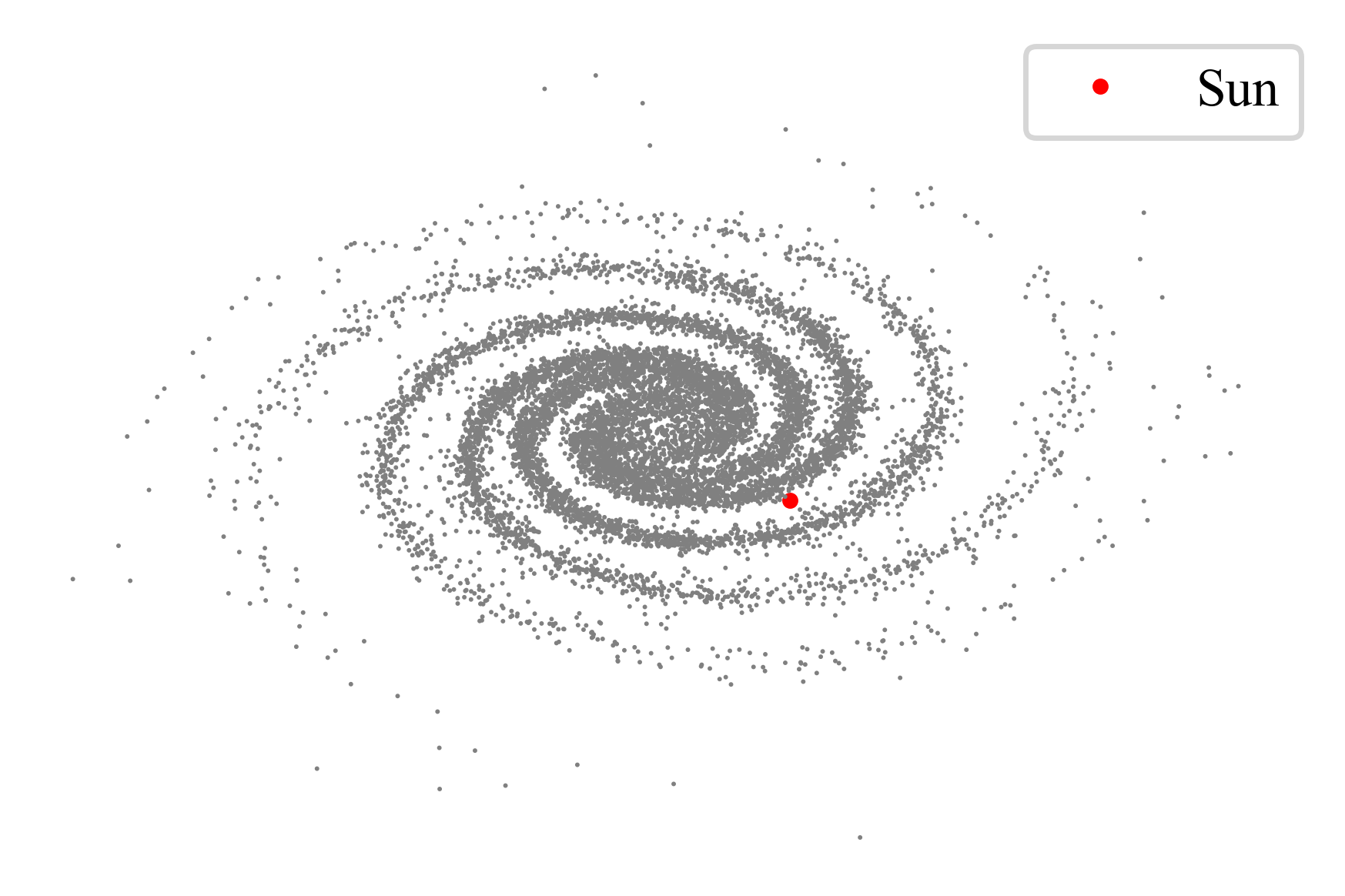}
    \caption{Initial location distribution of the newborn neutron stars using the models described above. The red point demonstrates the location of the solar system.
    \label{fig:birthloc}}
\end{figure}

The spatial distribution for newborn pulsars as described above is depicted in Figure~\ref{fig:birthloc}

\begin{equation}
    \rho(z)=\frac{1}{2z_0}\exp\left(-\frac{\left|z\right|}{z_0}\right).
\end{equation}

\subsubsection{Kick velocity}
Several direct observation evidence supports that neutron stars receive natal kicks during supernova explosions.
For instance, young pulsars can reach a velocity range from tens to over 1000\kms\ while their progenitors, i.e.\ O-B stars, have a low velocity of only $\sim$10--30\kms.
Moreover, the scale height of spatial distribution of pulsars in the Galaxy is $\sim$300\,pc \citep{lorimer2006parkes}, significantly larger than that of the progenitors ($\sim$50\,pc). 
Previous studies try different ways to parameterize the velocity distribution, including bimodal \citep{arzoumanian2002velocity}, single Maxwellian \citep{hobbs2005statistical}, and bimodal Maxwellian  \citep{verbunt2017observed,igoshev2020observed}.
In our simulation, we adopt the two-component isotropic Maxwellian distribution suggested by a recent work \citep{igoshev2020observed}
\begin{equation}
    f_{\rm max}(v\mid \sigma)=\sqrt{\frac{2}{\pi}} \frac{v^2}{\sigma ^3}
    \exp{\left(-\frac{v^2}{2\sigma^2}\right)}\,\mathrm{d}v \quad\mbox{and}
\end{equation}
\begin{equation}
    f(v\mid w,\sigma_1,\sigma_2)=wf_{\rm max}(v \mid \sigma_1)+(1-w)f_{\rm max}(v\mid \sigma_2).
\end{equation}
We fix $w=0.2, \sigma_1=56$\,km\,s$^{-1}$, and $\sigma_2=336$\,km\,s$^{-1}$,
obtained by fitting a sample of young pulsars with characteristic age younger than 3\,Myr old \citep{igoshev2020observed}.
The velocity is assumed to have a random direction drawn from an
isotropic distribution.

Note that we do not attempt to fit the pulsar spatial nor velocity distributions,
as they are insensitive to the magneto-rotational evolution of pulsars, which is
the focus of this work.

\subsubsection{Initial spin and magnetic field}
We regard the initial period $P_0$ and magnetic field $B_0$ as independent from other properties although they could be potential connection \citep[e.g., between $P_0$ and $v$; see][]{ng2007birth}.

In the previous population synthesis study, the pulsar initial magnetic field strength $B_0$ and period $P_0$ are assumed to draw from log-normal distributions \citep[e.g.,][]{2023arXiv231214848G}. 
This is supported by recent studies, which show that log-normal is preferred over normal distribution for both $B_0$ and $P_0$ \citep{igoshev2022initial,xu2023back}.

This is adopted in our simulation
\begin{eqnarray}
p(\log{P_0})&=&\frac{1}{\sqrt{2\pi\sigma_P^2}}\exp\left[-\frac{(\log P_0-\mu_P)^2}{2\sigma_P^2}\right]\\
p(\log{B_0})&=&\frac{1}{\sqrt{2\pi\sigma_B^2}}\exp\left[-\frac{(\log B_0-\mu_B)^2}{2\sigma_B^2}\right],
\end{eqnarray}
where $\mu_P$, $\sigma_P$, $\mu_B$, and $\sigma_B$, are free parameters determined by fitting.

Finally, for each simulated pulsar, we assigned a random initial inclination
angle $\alpha$ (between the magnetic and rotational axes) and a random viewing
angle $\xi$ (between the spin axis and line of sight), both drawn from isotropic
distribution in 3D, i.e.\ $\cos\alpha,\cos\xi\sim$ uniform(0,1).

\subsection{Pulsar evolution}

\subsubsection{Magneto-rotational evolution}
The spin-down process of pulsars converts its rotational kinetic energy into
electromagnetic radiation and particle outflow, resulting in an increasing period at a rate of \pdot.
For the simplest model of a rotating dipolar magnetic field in vacuum,
\pdot\ is related to field strength $B$ at the magnetic equator and the magnetic inclination angle $\alpha$
\begin{equation}
    B^2 =\frac{1}{\sin^2\alpha} \frac{3Ic^3}{8\pi^2R^6} P\dot P,
\end{equation}
where $I$ is the pulsar moment of inertia, $c$ is the speed of light, and $R$ is the pulsar radius.
Taking the canonical values of $R=10^6$\,cm and $I=10^{45}$\,g\,cm$^2$
gives
\begin{equation}
B=3.2\times10^{19}\sqrt{P\dot P/\sin^2\alpha}\,\mbox{G}.
\label{eqn:spindown0}
\end{equation}
While this standard spin-down formula was commonly adopted in previous population synthesis studies \citep[e.g.,][]{faucher2006birth,gonthier2007population,popov2010population}.
Recent numerical simulations of pulsar magnetosphere found a more 
realistic force-free spin-down model of
\begin{equation}
    \dot P=\frac{4\pi^2 R^6 B^2}{I c^3P}(k_0+k_1 \sin^2\alpha),
    \label{eqn:spindown1}
\end{equation}
i.e., $\sin^2\alpha$ dependence becomes $(k_0+k_1\sin^2\alpha)$, such that even an aligned rotator ($\alpha=0$) can spin down \citep{spitkovsky2006time,kalapotharakos2009three,petri2012pulsar,tchekhovskoy2013time,philippov2014time}.
During spin-down, the electromagnetic torque exerted by the plasma-filled magnetosphere
also makes the magnetic axis to progressively align with the rotational axis
\citep{beskin1988theory,philippov2014time,igoshev2020braking}.
We follow \citet{philippov2014time} to model this with
\begin{equation}
    \dot \alpha=-\frac{4\pi^2 R^6 B^2}{I c^3 P^2} k_2\sin\alpha\cos\alpha,
    \label{eqn:spindown2}
\end{equation}
where the numerical factor $k_0, k_1, k_2$ have a week dependence on the ratio between neutron star radius and light cylinder radius. For typical pulsar, $k_0\approx k_2\approx 1.0$, $k_1\approx 1.2$ \citep{philippov2014time}.
As mentioned in the introduction, there is observational evidence suggesting the decay of pulsar's magnetic field over its lifetime. Taking this into account in the population synthesis affects the evolution of \pdot\ and $\alpha$, and hence the
pulsar evolutionary tracks in the \ppdot\ diagram.
We consider in this work the $B$-field decay in the crust due to Ohmic loss and Hall effect.
The decay in the core is ignored since the core can be considered as a nearly neutral superconductor, which has a very long decay timescale.
Ohmic decay is the result of scattering between phonons and impurities of the crystalline lattice
\citep[e.g.,][]{pethick1995ohmic,igoshev2020braking}.
It can be parameterized by an exponential function
\begin{gather} 
    B(t)=B_0\exp{\left(-\frac{t}{\tau_{\rm Ohm}}\right)}, \label{eqn:spdexp}
\end{gather}
where timescale $\tau_{\rm Ohm}$ is related to the impurities, conductivity, and spatial scale of the crust \citep{igoshev2021evolution}.
On the other hand, Hall effect redistributes the magnetic energy, leading to non-linear decay as
\begin{equation}
    B(t)=\frac{B_0}{1+t/\tau_{\rm Hall}} \label{eqn:spdnon}
\end{equation}
with timescale $\tau_{\rm Hall}$ that depends on the initial magnetic field $B_0$ 

These two effects can be described by a more general form of $B$-field decay
\begin{gather}
    \frac{dB(t)}{dt}=-f_{B}B(t)^{1+\alpha_B}
\end{gather}
as suggested by numerical calculations \citep{colpi1999period,beniamini2019formation,jawor2022modelling}.
Here the power-law index $\alpha_B$ indicates the decay contribution of Ohmic dissipation and Hall effect.
Integrating the equation above, the magnetic field evolution thus follows:
\begin{equation}
    B(t)=\left\{
    \begin{array}{lr}
        B_0\exp{(-f_Bt)} &\mbox{if}\ \alpha_B=0 \\
        B_0(1+\alpha_{B}f_BB_0^{\alpha_B}t)^{-1/\alpha_B} &\mbox{if}\ \alpha_B>0.
    \end{array}
    \right.
    \label{eqn:bdecay}
\end{equation}
It is obvious that when $\alpha_B=0$, the equation above reduces to Eq.~\ref{eqn:spdexp} with $\tau_{\rm Ohm}=1/f_B$.
For the special case of $\alpha_B=1$, it reduces to Eq.~\ref{eqn:spdnon} with $\tau_{\rm Hall}=1/B_0f_B$.
In our simulations, we implement the two functions in Eq.~\ref{eqn:bdecay}
separately to avoid numerical instability when $\alpha_B$ is close to 0. We fit
$\tau_{\rm Ohm}$ for the exponential decay case and fit $f_B$ and
$\alpha_B$ for the power-law decay case. 
Finally, we also introduce a minimum magnetic field strength $B_{\rm min}=10^8$\,G \citep{zhang2006bottom}
so that it does not decrease to unphysical values.
We will show in Section~\ref{sec:result} that our result is insensitive to $B_{\rm min}$, since
almost no pulsar $B$-fields can reach this limit at the end of the simulation.


At each timestep of integration, the value of $B$ is first calculated, then used to determine changes of $P$ and $\alpha$
according to Eqs.~\ref{eqn:spindown1} and \ref{eqn:spindown2}.
The simulation of each pulsar is terminated at a maximum age randomly chosen within 100\,Myr.


\subsubsection{Dynamical evolution}
After a pulsar is born, we trace its orbit in the Galactic gravitational potential
during its lifetime as in previous studies \citep[e.g.,][]{cieslar2020markov}.
The gravitational potential mainly consists of contributions from the Galactic disk, bulge, and dark matter halo. We use the model given by \cite{miyamoto1975three} to characterize the gravitational potential of the bulge and the disc
\begin{eqnarray}
\Phi_{\rm bulge}&=&-\frac{GM_b}{\sqrt{b_b^2+R^2}}\\
\Phi_{\rm disc}&=&-\frac{GM_{d}}{\sqrt{\left(a_d+\sqrt{b_d^2+z^2}\right)^2+r_{xy}^2}},
\end{eqnarray}
respectively, where the bulge mass is $M_b=1.12\times10^{10}\,M_\odot$ and disc mass is $M_d=8.78\times10^{10}\,M_\odot$, $b_b=0.277$\,kpc, $a_d=4.2$\,kpc, and $b_d=0.198$\,kpc. $R$ and $r_{xy}$ are the distance and radial distance to the Galactic Center, respectively.
For the halo potential, we follow the model by \citet{1990ApJ...348..485P}:
\begin{equation}
    \Phi_{\rm halo}=-\frac{GM_h}{2r_c}\left[\frac{2r_c}{R} \arctan\left(\frac{R}{r_c}\right)+\ln\left(1+\frac{r_c^2}{R^2}\right)\right],
\end{equation}
where the halo mass is $M_h=5\times10^{10}\,M_\odot$ and $r_c=6$\,kpc.
Since the mass of halo is divergent, we applied a cutoff radius of $r_{cut}$=100\,kpc, beyond which the halo density drops to zero and $\Phi_{\rm halo}\varpropto r^{-1}$, following \citet{belczynski2010double}.


\subsection{Pulsar Emission}
After simulation, we calculate the radio and Gamma-ray emission of a pulsar
using their final properties (e.g., $P$, $B$, viewing angle, distance, DM, etc.)
to determine if they are detectable by the surveys.

\subsubsection{Radio Emission}
The observed pulsar radio emission geometry can be modeled by core and cone components \citep{rankin1993toward}.
This empirical model has been widely used in previous population synthesis studies \citep[e.g.,][]{arzoumanian2002velocity, gonthier2004role,harding2007geminga,gonthier2018population}. In this work, we follow \citet{gonthier2018population} to describe the beam geometry and radio flux density $S_{\nu}$.

First, we adopt an empirical power law function to parameterize the total radio luminosity
\begin{equation}
    L_{\nu}=L_{\rm cone}+L_{\rm core}=f_{\nu} P^{\alpha_{\nu}}_{-3} {\dot{P}}^{\beta_{\nu}}_{-21},
    \label{eqn:radiolum}
\end{equation}
where the total radio luminosity $L_\nu$, the cone luminosity $L_{\rm cone}$ and the core luminosity $L_{\rm core}$ are measured in mJy\,kpc$^2$\,MHz, $P_{-3}$ is the spin period in $10^{-3}$\,s, ${\dot P}_{-21}$ is the spin-down rate in $10^{-21}$\,s/s, $f_{\nu}$, $\alpha_{\nu}$ and $\beta_{\nu}$ are free parameters determined by fitting.
The ratio $\mathcal{R}$ between the core and cone luminosities depends on $P$, $\dot P $ with a broken power law,
\begin{equation}
    \mathcal{R}=\left\{
    \begin{array}{lr}
        6.2\ P\ \dot{P}_{-15}^{-0.07} & P < 0.7\,\mbox{s} \\
        P^{-2.1}\ \dot{P}^{-0.07}_{-15} & P \ge 0.7\,\mbox{s}.
    \end{array}
    \right.
\end{equation}
The simulated pulse profile of radio emission is axisymmetric and depends on the survey frequency $\nu$ and polar angle $\theta$ as
\begin{equation}
    F(\nu,\theta)=F_{\rm core}\ e^{-{\theta^2}/{\rho_{\rm core}}^2}+F_{\rm cone}\ e^{-(\theta-\bar{\theta})^2/{w_e}^2},
\end{equation}
where
\begin{equation}
    F_i(\nu)=\frac{-1+\alpha_i}{\nu}\left(\frac{\nu}{50\mbox{\,MHz}}\right)^{\alpha_i+1}\ \frac{L_i}{\Omega_i D^2}.
    \label{eqn:flux}
\end{equation}
$\rho_{\rm core}$ and $w_e$ are the effective width of core and cone beam, $\bar \theta$ is the hollow cone beam annulus size. 
The index $i$ refers to the cone or core component. The spectral index $\alpha_i$ is given by \cite{harding2007geminga} and $D$ is the distance in kpc. The polar angle $\theta$ is related to the magnetic inclination angle $\alpha$, the viewing angle $\xi$, and the rotational phase $\phi$ by
\begin{equation}
\cos{\theta}=\sin{\xi}\sin{\alpha}\cos{\phi}+\cos{\alpha}\cos{\xi}.
\label{eqn:polar}
\end{equation}

\begin{figure}[tbh]
\centering
\includegraphics[width=0.45\textwidth]{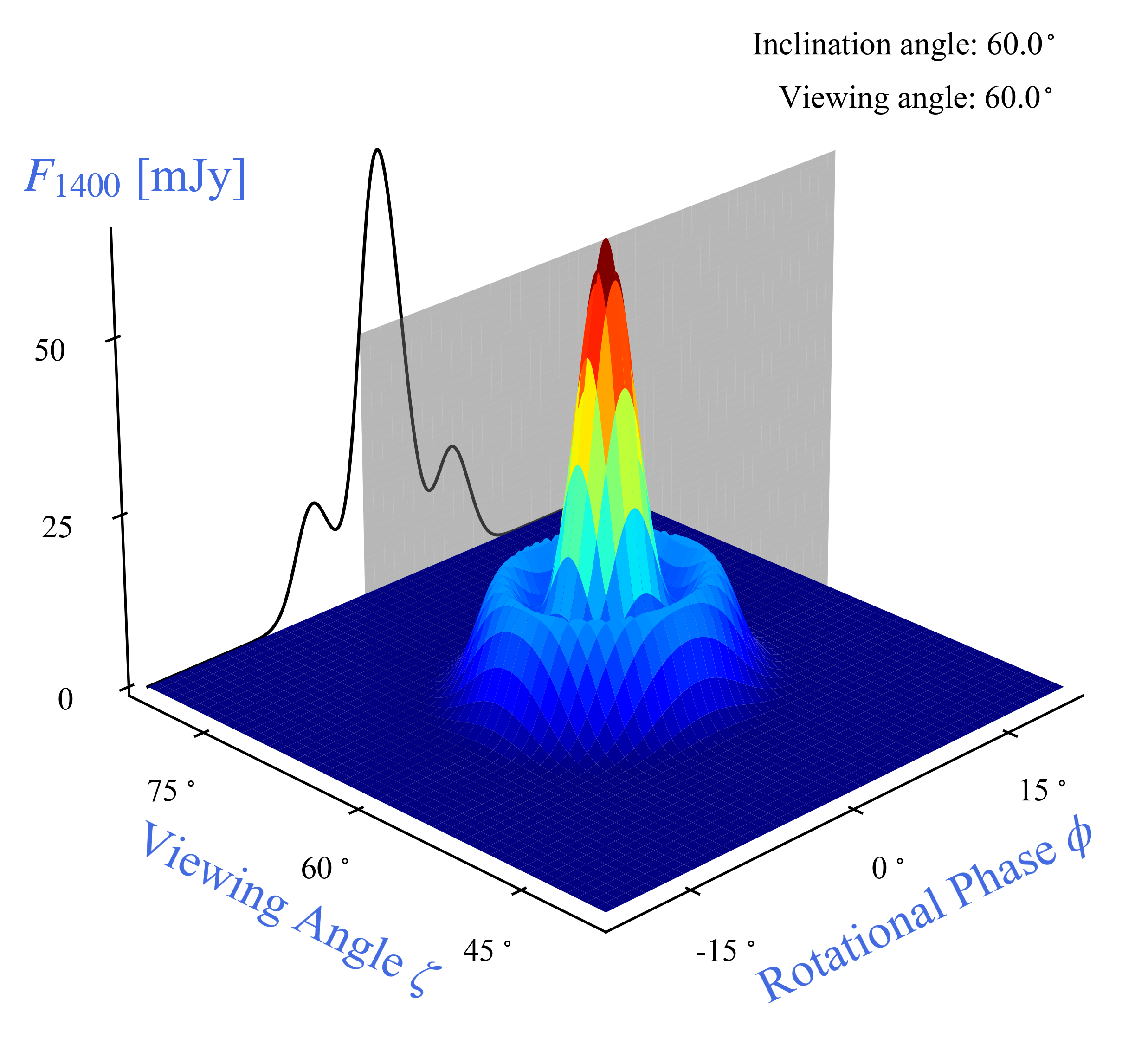}
\caption{Radio beam of cone and core model. The projected pulse profile changes with different viewing angles and inclination angles. The
pulsar properties are taken with typical values in this example ($\alpha$ = $60^\circ$, $\xi$ = $60^\circ$, $P = 0.3$\,s, $\pdot=10^{-14}$\,s/s and $B = 10^{12}$\,G). $F_{1400}$ represents the
flux density in unit of mJy at 1400\,MHz.
\label{fig:beam}}
\end{figure}

The normalization factors are given by
\begin{equation}
    \begin{aligned}
        &\Omega_{\rm core}=\pi\rho_{\rm core}^2\\
        &\Omega_{\rm cone}=2{\pi}^{3/2} {\omega_e} \bar{\theta}.
    \end{aligned}
\end{equation}
The core beam is a Gaussian function that scales with spin period as
\begin{equation}
    \rho_{\rm core}=1.5^\circ P^{-0.5}.
\end{equation}
The cone beam has a hollow Gaussian form with the annulus size and width as
\begin{equation}
    \begin{aligned}
        &\bar{\theta}=(1-2.63 \delta_w)\rho_{\rm cone}\\
        &w_e=\delta_w \rho_{\rm cone},
    \end{aligned}
\end{equation}
respectively, where $\delta_w=0.18$ \citep{gonthier2006developing} and the opening angle 
depends on the pulsar emission altitude $r_{\rm KG}$ and spin period $P$ \citep{kijak1998radio} as
\begin{equation}
    \rho_{\rm cone}=1.24^{\circ} r_{\rm KG}^{0.5} P^{-0.5},
\end{equation}
with $r_{\rm KG}$ estimated using the pulse width method \citep{kijak2003radio}
\begin{equation}
r_{\rm KG}=40\dot{P}^{0.07}_{-15}P^{0.3}\nu_{\rm GHz}^{-0.26},
\end{equation}
where $\nu_{\rm GHz}$ is the observation frequency in GHz.
Eq.~\ref{eqn:flux} shows how radio flux density drops as the polar
angle $\theta$ increases. Given $\xi$ and $\alpha$, the observed pulse intensity
variation with the rotational phase $\phi$ can be calculated according to Eq.~\ref{eqn:polar}.
This gives the radio pulse profile, see Figure~\ref{fig:beam} for an example. We then calculate the phase-averaged radio flux density to determine if
the pulse is brighter than the minimum detectable threshold of different radio
surveys.

We follow some previous studies not to assume a death line for the radio emission
\citep[e.g.,][]{gullon2014population,2023arXiv231214848G}, such that the
intrinsic radio emission of old pulsars does not completely shut down. On the
other hand, the assumed luminosity law (Eq.~\ref{eqn:radiolum}) will naturally
make old pulsars become too faint to be detected.

\subsubsection{Gamma-ray luminosity and beaming geometry \label{sec:gammaflux}}
One of the applications of our population synthesis study is making predictions for the number of
detectable Gamma-ray pulsars.
We model the Gamma-ray luminosity with the fundamental plane relation 
\citep{kalapotharakos2022fundamental}
\begin{equation}
    L_{\gamma}=10^{14.3} \varepsilon_{\rm cut}^{1.39} B_p^{0.12} \dot E^{0.39},
\end{equation}
where $\varepsilon_{\rm cut}$ is the cutoff energy in units of MeV obtained from the empirical $\varepsilon_{\rm cut}$-$\dot E$ relations \citep{kalapotharakos2017fermi},
$B_P$ is the field strength at the magnetic pole, i.e.\ twice the value of $B$ in our simulation,
and $L_\gamma$ and $\dot E$ are in units of \ergs.
The observed phase-averaged Gamma-ray flux is given by
\begin{equation}
    F_\gamma=\frac{L_\gamma}{4\pi f_\Omega D^2},
\end{equation}
where the beaming factor $f_\Omega$ depends on the viewing angle $\xi$, $\alpha$, and Gamma-ray
efficiency ($w=$)$\eta$ ($\propto \dot E^{1/2}$).

We adopt the results from the outer gap (OG) and two-pole
caustic (TPC) models to estimate $f_\Omega$ \citep{watters2009atlas}.
For the OG model,
\begin{equation}
    f_{\Omega}\approx\left\{ \begin{array}{lr}
        0.17-0.69w+(1.15-1.05w)(\alpha /90^\circ)^{1.9} & \xi<60^\circ \\
        0.17-0.69w+(1.15-1.05w)(\alpha /90^\circ)^{1.9}\\
        -1.35(2/3-\xi/90^\circ) & \xi>60^\circ.
    \end{array} \right.
\end{equation}
For the TPC model,
\begin{equation}
    f_{\Omega}\approx\left\{ \begin{array}{lr}
        0.8+1.2(0.3-w)\cos(2\beta) & \xi>\xi_I \\
        0.3+1.5(1-w)[1+(\xi-\xi_I)/90^\circ] & \xi<\xi_I,
    \end{array} \right.
\end{equation}
where $\beta=\xi-\alpha$ and $\xi_I$ relates to the boundary of hollow cone above the null charge surface
\begin{equation}
    \xi_I=(75+100w)-(60+1/w)(\alpha/90^\circ)^{2(1-w)}.
\end{equation}
We do not attempt to fit any of the parameters here due to the relatively small
sample size of Gamma-ray pulsars. Therefore, this has no impact on other
parameters of our population model and $B$-field evolution.

\subsection{Radio Detection}
To determine the visibility of a simulated pulsar for a specific radio survey,
we first check if the pulsar's final location lies inside the survey region.
If so, we calculate the dispersion measure (DM) from its final location
with the latest free electron density model of the Galaxy \citep{yao2017new}.
Then we employ the standard radiometer formula \citep{dewey1985search} to compare
the pulsar flux density and the detection threshold $S_{\rm min}$ of the survey
\begin{equation}
    S_{\rm min}=\frac{C\beta [T_{\rm rec}+T_{\rm sky}(l,b)]}{G \sqrt{N_p t_{\rm obs} \Delta F}}\sqrt{\frac{W_e}{P-W_e}},
\end{equation}
where $C$ is the detection signal-to-noise threshold, $\beta$ is the degradation factor
caused by system loss, $T_{\rm rec}$\, is the receiver system temperature in K, $T_{\rm sky}(l,b)$ is the sky background temperature in K obtained from the all-sky atlas \citep{haslam1982408}, $G$ is the effective
gain of the telescope antenna in units of K\,Jy$^{-1}$,
$N_p$ is the number of polarization, $t_{\rm obs}$ is the observation time in s,
$\Delta F$ is the total bandwidth in MHz, $P$ is the pulsar period in s, and
$W_e$ is the effective pulse width in ms.
We estimate $W_e$ based on the formulas given by \citet{bhat2004multifrequency}
and we follow previous studies \citep[e.g.,][]{lorimer1993pulsar,bates2014psrpoppy} to estimate
$G$ from the original telescope gain $G_0$ listed in Table~\ref{tab:surveypars} by
\begin{equation}
     G=G_0\exp\left({\frac{-2.77r^2}{w^2}}\right).  \label{eq:beam}
\end{equation}
This accounts for the decrease in sensitivity when a pulsar has an offset $r$ from
the telescope pointing center. Here $w$ is the full width at half maximum (FWHM)
of the telescope beam and $r$ is randomly chosen assuming uniform distribution
of $r^2$ between 0 and $w^2/4$ \citep{bates2014psrpoppy}. 



If a simulated pulsar lies within the survey area with radio flux density above the
survey sensitivity, then it is considered as detected and its parameters will be compared with
the observed pulsar sample using the procedure described in Section~\ref{obscomp} below.
Regardless of whether a simulated pulsar is detected or not, it is always counted for the birth rate calculation.

\begin{deluxetable*}{ccccccccccccccc}[ht] 
\tablecaption{Number of normal pulsars detected from radio surveys compared with our best-fit model simulation results. \label{tab:surveys}}
\tablehead{\colhead{Survey\tablenotemark{a}} & \colhead{ar1} & \colhead{ar3} & \colhead{ar4} & \colhead{palfa} & \colhead{gb2} & \colhead{gb3} & \colhead{gbncc} & \colhead{htru\_pks} & \colhead{mol2} & \colhead{pkshl} & \colhead{pks70} & \colhead{pkssw} & \colhead{pksmb} & \colhead{lotaas}}
\startdata
Obs.\ no. & 48   & 59   & 103   & 249   & 75   & 86    & 307   & 981    & 220   & 31   & 271   & 196   & 1016  &257 \vspace*{5pt}\\
Sim.\ no. & 49 & 68 & 166 & 255 & 81 & 164 & 467 & 1226 & 256 & 31 & 204 & 195 & 729 & 310\vspace*{2pt}
\enddata
\tablenotetext{a}{See Table~\ref{tab:surveypars} in Appendix~\ref{appendix:a} for the full name of the surveys.}
\end{deluxetable*}

\subsection{Model Comparison and Parameter Optimization \label{obscomp}}
\subsubsection{Pulsar sample \label{obssample}}
We select our pulsar sample from major surveys listed in the ATNF catalog\footnote{\url{https://www.atnf.csiro.au/research/pulsar/psrcat/index.html}} \citep{manchester2005australia} with the following criteria:
\begin{enumerate}
\item $P>0.03$\,s and $\dot P >0$ to rule out millisecond pulsars.
\item Exclude rotating radio transients (RRATs) and binary pulsars, as these could have different formation channels and evolution paths than normal pulsars. 
\item Exclude radio surveys with fewer than 25 normal pulsars to speed up the computing time.
These are listed in Table~\ref{tab:nosurvey} in Appendix~\ref{appendix:a}. This only leads to 38 fewer pulsars, not significant.

\item Exclude surveys that are fully overlapped with others (see Table~\ref{tab:nosurvey} in Appendix~\ref{appendix:a}).

\item Exclude ongoing surveys (e.g., the MeerKAT TRAPUM Survey).

\item Exclude 87 pulsars that have DM-estimated distance larger than 25\,kpc, which are likely to be extra-galactic.
\end{enumerate}
The above selection criteria give a total of 1859 radio pulsars from 14 surveys.
covering the vast majority of normal radio pulsars.
Table~\ref{tab:surveys} shows the number of pulsars from each survey and the survey parameters
are presented in Table~\ref{tab:surveypars}.


Our simulations assume that the radio flux density is measured at 1400\,MHz ($S_{1400}$).
However, there are 235 pulsars in our sample that only have measurements in other bands.
We scale their flux densities to 1400\,MHz assuming a typical spectral index of $-1.60$
\citep{jankowski2018spectral}. We show in Appendix~\ref{appendix:a} that this does not
change the flux density distribution.

\subsubsection{Fitting of the model parameters}
To evaluate the goodness of the model fit given a set of parameters,
we compare the distribution of simulated and observed pulsar properties using multivariate
KS statistic \citep{fasano1987multid,justel1997multivariate}.
The procedure is outlined in Appendix~\ref{Appendix:B}. We follow previous works
\citep[e.g.,][]{gonthier2018population} to focus on four observables: $P$, \pdot,
DM, and $S_{1400}$ as other observables are not sensitive to the
magneto-rotational evolution model that we are interested in.


The fitting parameters are $\mu_P$, $\sigma_P$, $\mu_B$, $\sigma_B$, $\alpha_B$, $f_B$, $f_\nu$, 
$\alpha_\nu$, and $\beta_\nu$ for the power-law decay model. For the Ohmic decay model, $f_B$ is
replaced with $\tau_{\rm Ohm}$ and there is one fewer free parameter as $\alpha_B=0$.
We employed the MCMC technique to search for the optimal model parameters.
Given a KS statistic $D_{\rm ks}$, the $p$-value can be calculated as
\begin{equation}
    \begin{aligned}
    P(>Z)&=2\sum_{k=1}^{\infty}(-1)^{k-1}e^{-2k^2Z^2}\\
    Z&=\sqrt{\frac{n_1n_2}{n_1+n_2}}D_{\rm ks},
    \end{aligned}
\end{equation}
where $n_1$ and $n_2$ are the sample size of simulation and observation
\citep{fasano1987multid}.
This is then is used to construct the likelihood function in MCMC. There are no truncated values for the free parameters. We assume log-uniform priors for $\tau_{\rm Ohm}$, $f_{\nu}$ and uniform priors for all other fitting parameters,
and employed the \texttt{emcee}
package \citep{foreman2013emcee} to run the MCMC chains.

\begin{figure*}
    \centering
    \includegraphics[width=\textwidth]{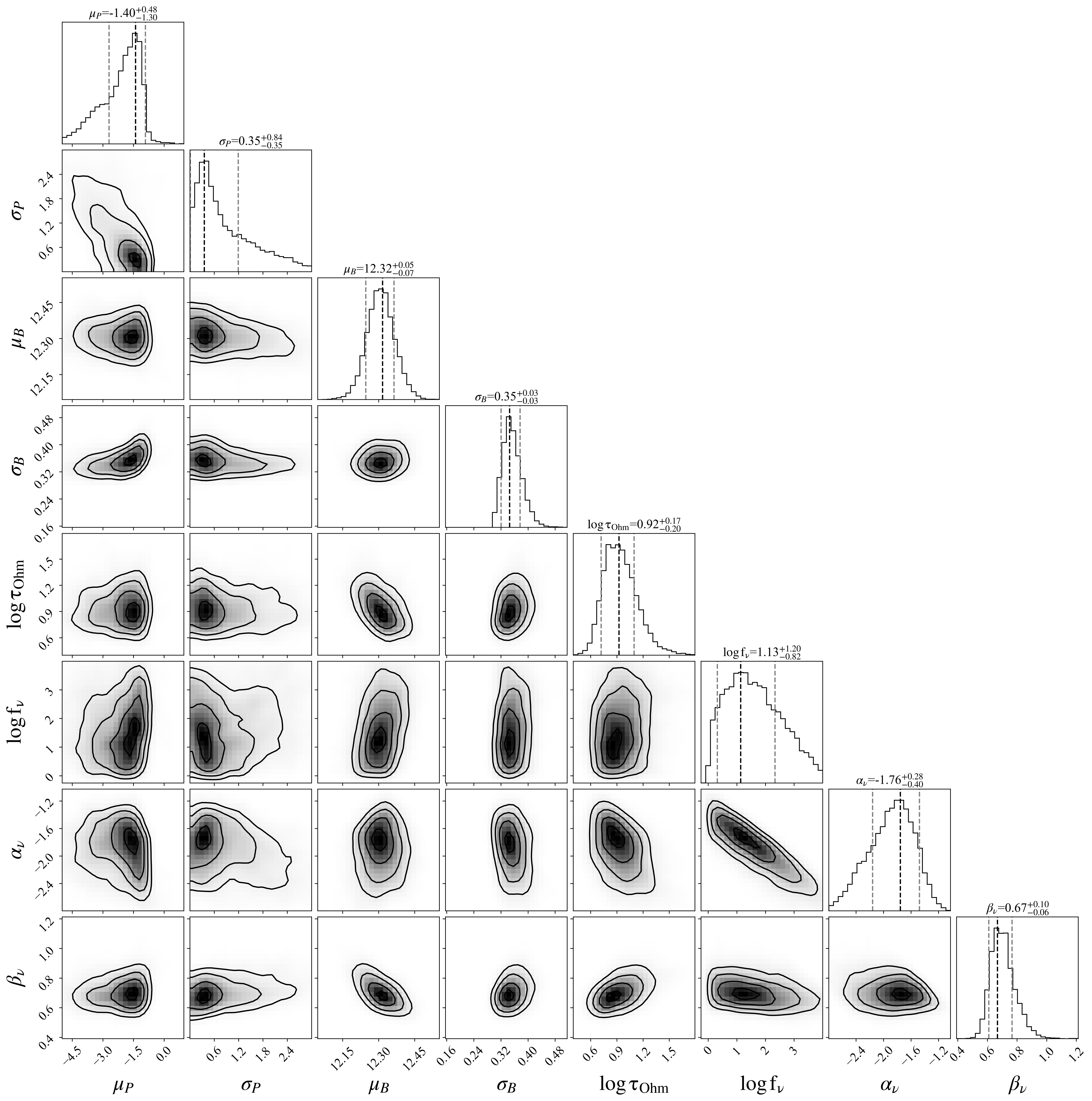}
    \caption{The MCMC marginal parameter space for Ohmic dissipation assumption. The 2D marginal plot is smoothed using a Gaussian kernel.}
    \label{fig:MCMC_Ohm}
\end{figure*}

\section{Results \label{sec:result}}
\subsection{Best-fit model parameters}
The best-fit power-law model suggests a very small value of $\alpha_B$ close to 0, which reduces the model to exponential decay. As we will show in Sect.~\ref{ch:tracks} below, the model with large $\alpha_B$ produce pulsars narrowly distributed in \ppdot\ space, which does not fit the observed pulsar population.

\begin{deluxetable*}{ccccccccccc}
    \tablecaption{Best-fit parameters with 1$\sigma$ confidence intervals for the exponential and power-law $B$-field decay models.\label{tab:pars}}
    \tablehead{\colhead{$\mu_P$} & \colhead{$\sigma_P$} & \colhead{$\mu_B$} &  \colhead{$\sigma_B$} &
    \colhead{$\log\tau_{\rm Ohm}$ (Myr)} &  \colhead{$\alpha_B$} & \colhead{$\log f_B$} &
    \colhead{$\log f_\nu$} & \colhead{$\alpha_\nu$} & \colhead{$\beta_\nu$} & \colhead{$D_{\rm ks}$}}
    \startdata
    \multicolumn{5}{l}{Exponential decay}\\
    $-1.40^{+0.48}_{-1.30}$ & $0.35^{+0.84}_{-0.35}$ & $12.32^{+0.05}_{-0.07}$ & $0.35\pm0.03$ &$0.92^{+0.17}_{-0.20}$ &
    \nodata & \nodata & $1.1^{+1.2}_{-0.8}$ & $-1.76^{+0.28}_{-0.40}$ & $0.67^{+0.10}_{-0.06}$ & $0.080$ \vspace*{5pt}\\
    \multicolumn{5}{l}{Power-law decay}\\
    $-1.62^{+0.3}_{-0.6}$ & $0.36^{+0.41}_{-0.18}$ & $12.33^{+0.04}_{-0.07}$ & $0.35^{+0.04}_{-0.05}$ &
    \nodata & $0.02^{+0.08}_{-0.02}$ & $-15.6^{+0.3}_{-0.7}$ &$0.9^{+0.3}_{-0.2}$ & $-1.45^{+0.08}_{-0.18}$ & $0.68^{+0.07}_{-0.05}$ & $0.082$ \vspace*{10pt}
    \enddata
\end{deluxetable*}

Figure~\ref{fig:MCMC_Ohm} shows the corner plot for the exponential model parameters, obtained from MCMC with 4 independent chains of 16 walkers and 2500 steps each, a total of 160,000 simulations. The optimal parameters determined from the most probable values (MPVs), i.e.\ at the maximum marginal probability density, with 68\% confidence intervals, are listed in Table~\ref{tab:pars}.
As indicated in the figure, the model parameters $\mu_B$, $\sigma_B$, $\tau_{\rm Ohm}$ and $\beta_\nu$ are well constrained, but $f_\nu$, the lower bound of $\mu_P$ and upper bound of $\sigma_P$ are not very well determined.
In addition, there exists strong correlations between $f_\nu$ and $\alpha_\nu$, $\mu_P$ and $\sigma_P$, and
some correlations among $\tau_{\rm Ohm}$, $\mu_B$, $\alpha_\nu$, and $\beta_\nu$.

\begin{figure*}[hbt!]
    \centering
    \includegraphics[width=\textwidth]{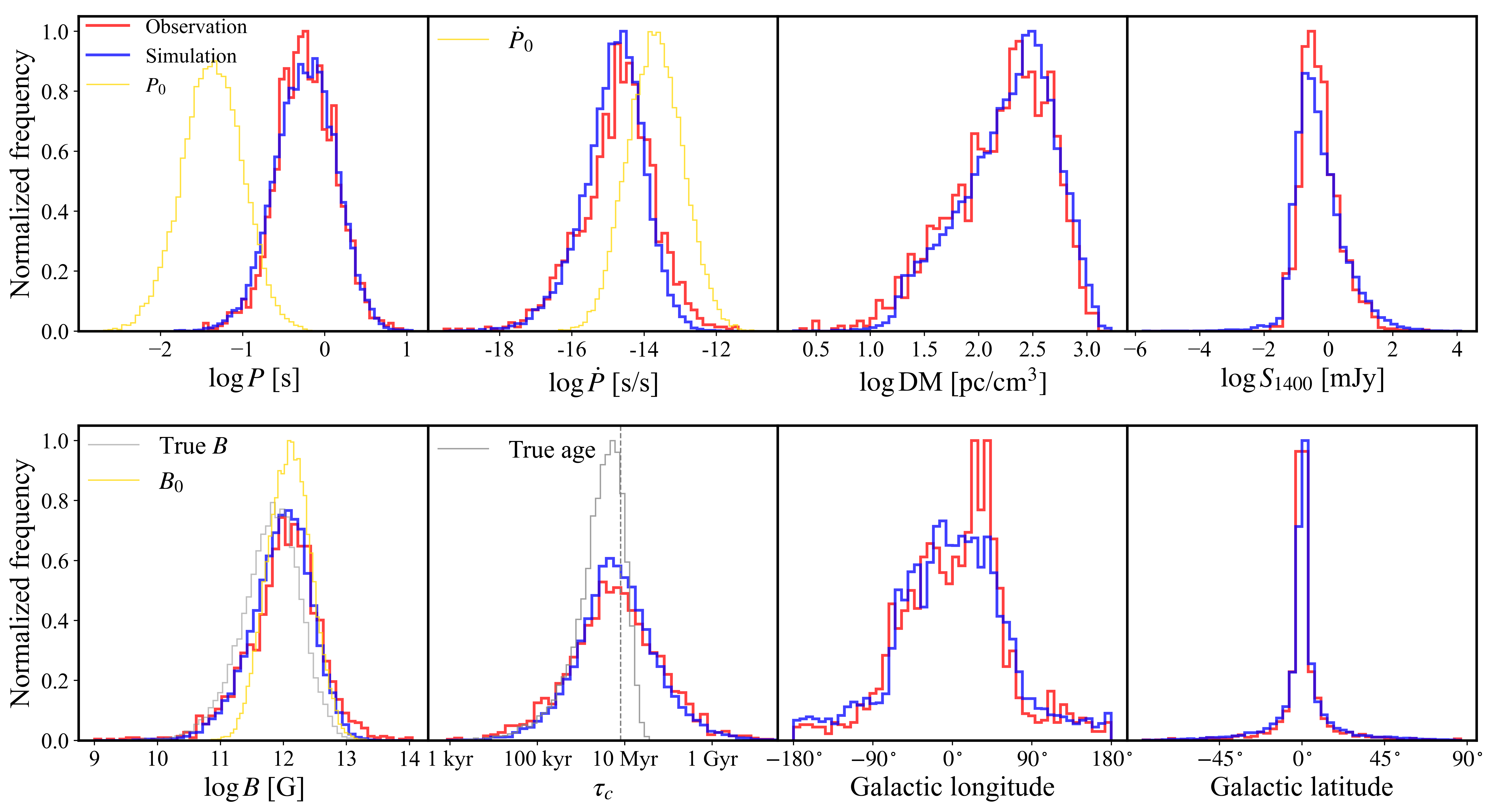}
    \caption{Distributions of observables for the simulated (blue) and observed (red) pulsar populations.
    The simulation is based on the exponential $B$-field decay model with the best-fit parameters.
    The initial spin period ($P_0$), initial spin-down rate ($\dot P_0$) and birth magnetic field strength ($B_0$)
    distributions (yellow histograms) are also shown for comparison. The magnetic field distributions shown
    by the red and blue lines are $B_{\rm SD}$ from Eq.~\ref{eqn:Bppdot}, and the true $B$-field from the simulations is plotted in grey color.
    The characteristic age $\tau_c$ is calculated with $P/2\dot P$, and the grey line shows the true age from simulations.
    The vertical dashed line in the plot indicates the best-fit Ohmic decay timescale of 8.3\,Myr.
    \label{fig:Ohm_hist}}
\end{figure*}

\subsection{Best-fit pulsar population}
Table~\ref{tab:surveys} shows a comparison between the observed and simulated number of pulsars in different surveys. The latter is obtained from the average of 100 simulations using the best-fit model parameter. 
These numbers are generally consistent and are very well matched for the palfa and pkshl surveys. However, some surveys, including ar4, gb3, and pksmb, show a significant discrepancy of up to 50\%. The possible cause of this will be discussed in Section~\ref{predict}.
In Figure~\ref{fig:Ohm_hist} we compare the distributions of observables for the simulation and pulsar sample.
We plot the four observables in the first row of Figure~\ref{fig:Ohm_hist} ($P$,
\pdot, DM, and $S_{1400}$) that are used to determine the goodness-of-fit, and
also the pulsar Galactic coordinates, the $B$-field, and the characteristic age
$\tau_c\equiv P/2\dot P$. The results show that our model can generally
reproduce the observed pulsar population. The only exception may be the Galactic
longitude, which has obvious excess at $l\sim30\arcdeg$ that our model fails to
capture. This problem was first noted by \citet{faucher2006birth} and is due to
the spiral arm structure that cannot be accurately reproduced by the simple
spatial model in Eq.~\ref{eqn:arm}.

\begin{figure*}[htb!]
    \centering
    \includegraphics[width=0.80\textwidth]{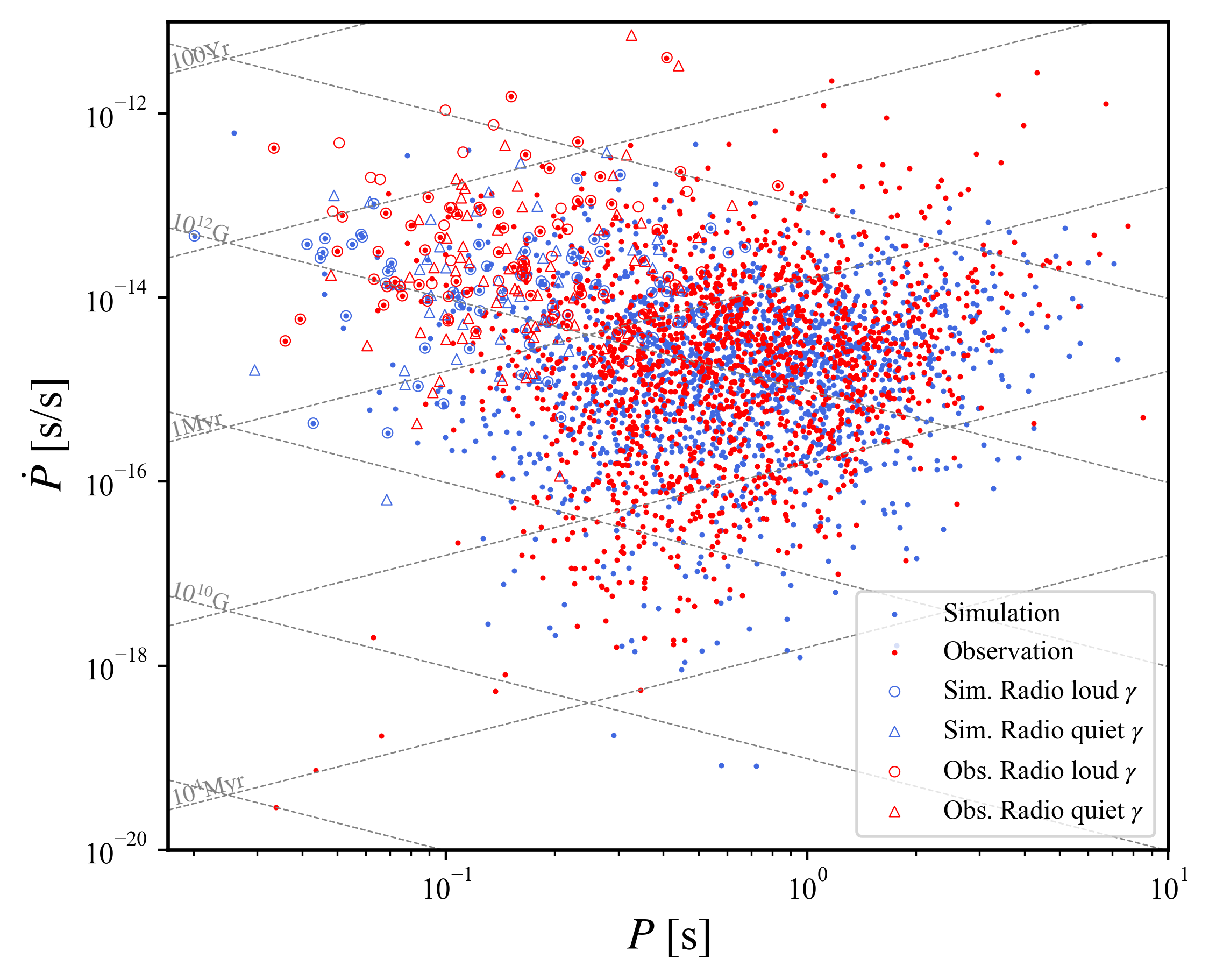}
    \caption{\ppdot\ diagram for the observed (red) and simulated (blue) pulsars.
    The simulation is based on the exponential $B$-field decay model with the best-fit parameters.
    The red dots show the pulsar sample from radio surveys listed in Table \ref{tab:surveypars}.
    The red open circles and triangles show radio-loud and radio-quiet Gamma-ray pulsars, respectively, from the Third Fermi-LAT catalog \citep{smith2023third}. 
    \label{fig:ppdot}}
\end{figure*}

\begin{figure*}[thb!]
    \centering
    \includegraphics[width=\textwidth]{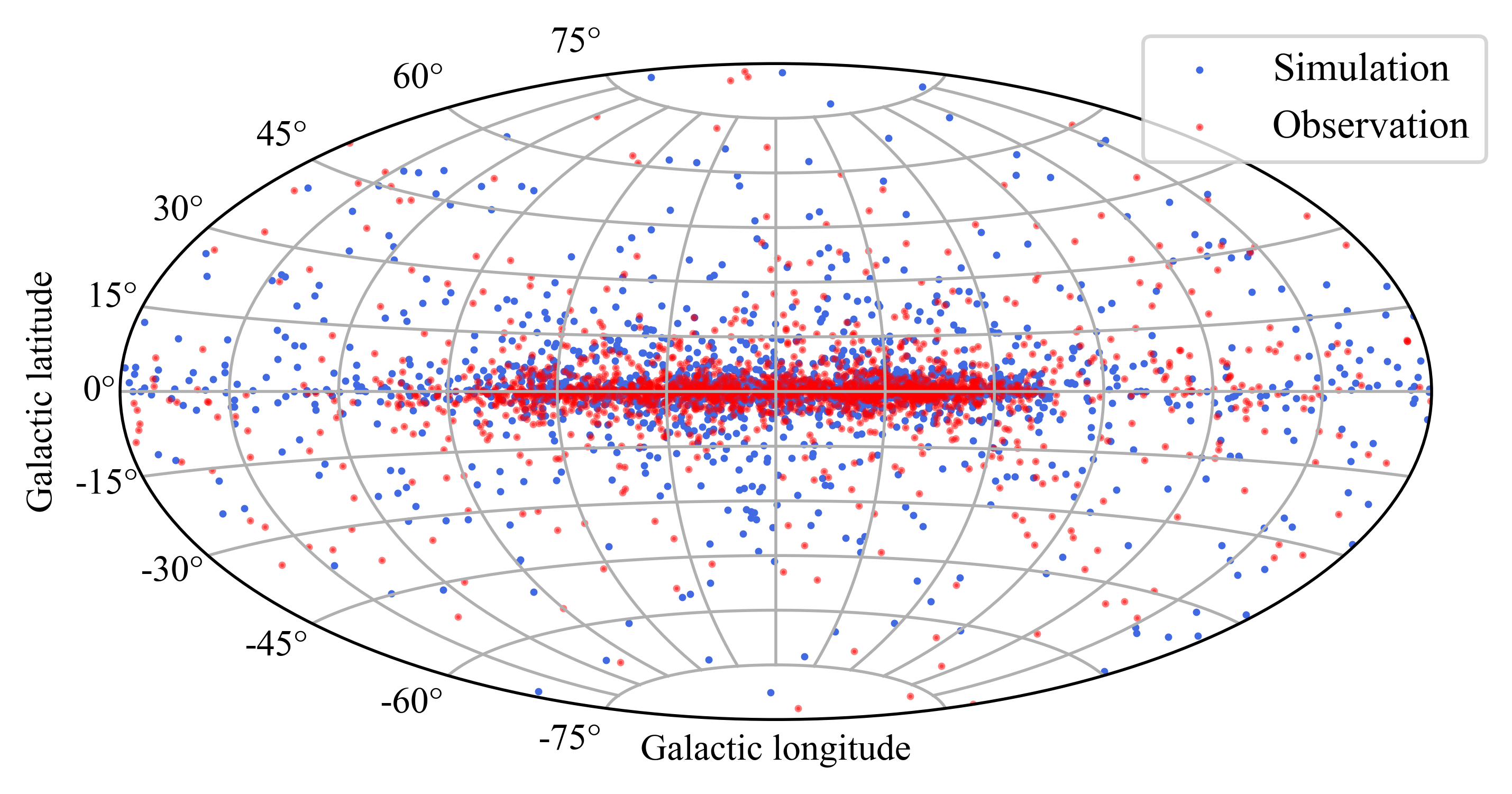}
    \caption{Galactic location distributions for the simulated (blue) and observed (red) pulsar populations.
    The simulation is based on the exponential $B$-field decay model with the best-fit parameters.
    \label{fig:Galactic}}
\end{figure*}

\begin{figure*}[bht!]
    \centering
    \includegraphics[width=0.8\textwidth]{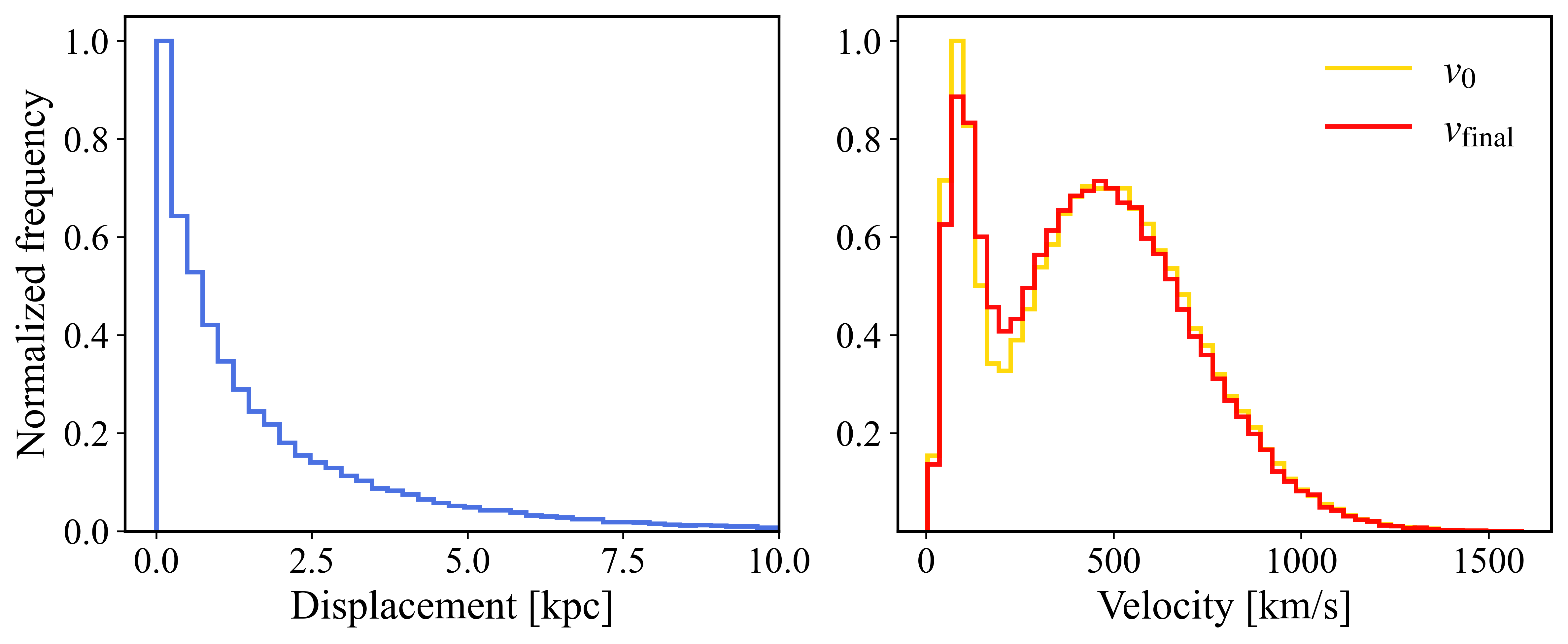}
    \caption{Dynamical evolution of simulated pulsars that are detected by the surveys.
    \emph{Left:} distribution of pulsar displacement between their birth sites and final locations.    
    \emph{Right:} distribution of initial and final pulsar velocities at the end of the simulation.
    \label{fig:Dyna}}
\end{figure*}

We also show in the figure a comparison between the true $B$-field in the simulation and the $B$-field estimated from spin down using Eq.~\ref{eqn:spindown0} with $\sin\alpha=1$, i.e.
\begin{equation}
    B_{\rm SD}=3.2\times10^{19}\sqrt{P\dot P}\,\mbox{G}, \label{eqn:Bppdot} 
\end{equation}
which is commonly adopted in the literature. They have very similar distributions and the true $B$-field is slightly smaller.
This is expected as a direct comparison between Eqs.~\ref{eqn:spindown1} and \ref{eqn:Bppdot} gives $B/B_{\rm SD}=\sqrt{3/(2+2.4\sin^2\alpha)}$, which varies between 0.55 and
0.82. In other words, Eq.~\ref{eqn:Bppdot} is indeed a good approximation to the realistic force-free spin-down model. Finally, we show the age distribution of the simulated pulsars and its comparison with $\tau_c$. The mean true age is 4.8\,Myr and over 40\% of them are older than $\tau_{\rm Ohm}$.
$\tau_c$ shows a good agreement with the true age for young pulsars up to $\sim1$\,Myr, and then becomes much larger beyond that. This can be attributed to the decay of $B$-field, which makes \pdot\ to drop rapidly (see discussion in Sect.~\ref{ch:tracks} below), thus, resulting in large $\tau_c$.

Figure~\ref{fig:ppdot} shows the \ppdot\ diagram of the observed pulsar sample and a random realization of the best-fit model with the same number of pulsars. They generally show a good match. We notice that our model fails to reproduce pulsars in the lower left corner of the diagram (i.e.\ with fast spin but small
\pdot). These could be mildly recycled pulsars and their formation is not modeled in our simulation.

\subsection{Dynamical evolution}
We plot in Figure~\ref{fig:Galactic} the location of the pulsar sample in Galactic coordinates
and a Monte Carlo realization of our model for comparison. They show a good
agreement in general. Results of pulsar dynamical evolution are plotted in
Figure~\ref{fig:Dyna}. These include distributions of pulsar displacements and
velocities. The pulsar displacement plot suggests that most pulsars do not
travel far over their lifetime. Over 78\% of them are found within 3\,kpc of
their birthsite at the end of the simulation. Similarly, the velocity
distribution plot shows that the final velocities ($v_{\rm final}$) do not
change much from the initial values ($v_0$). They follow the same double
Mawellian distribution with identical peak locations, but slightly fewer
($\sim$1.3\% drop) pulsars in the low-velocity component ($<130$\,\kms) and more
($\sim$1.1\% increase) with 200$-$300\kms. All these results suggest that acceleration due to Galactic gravitational potential is not significant for most of pulsars.

\begin{figure*}[htb]
    \centering
    \includegraphics[width=0.7\textwidth]{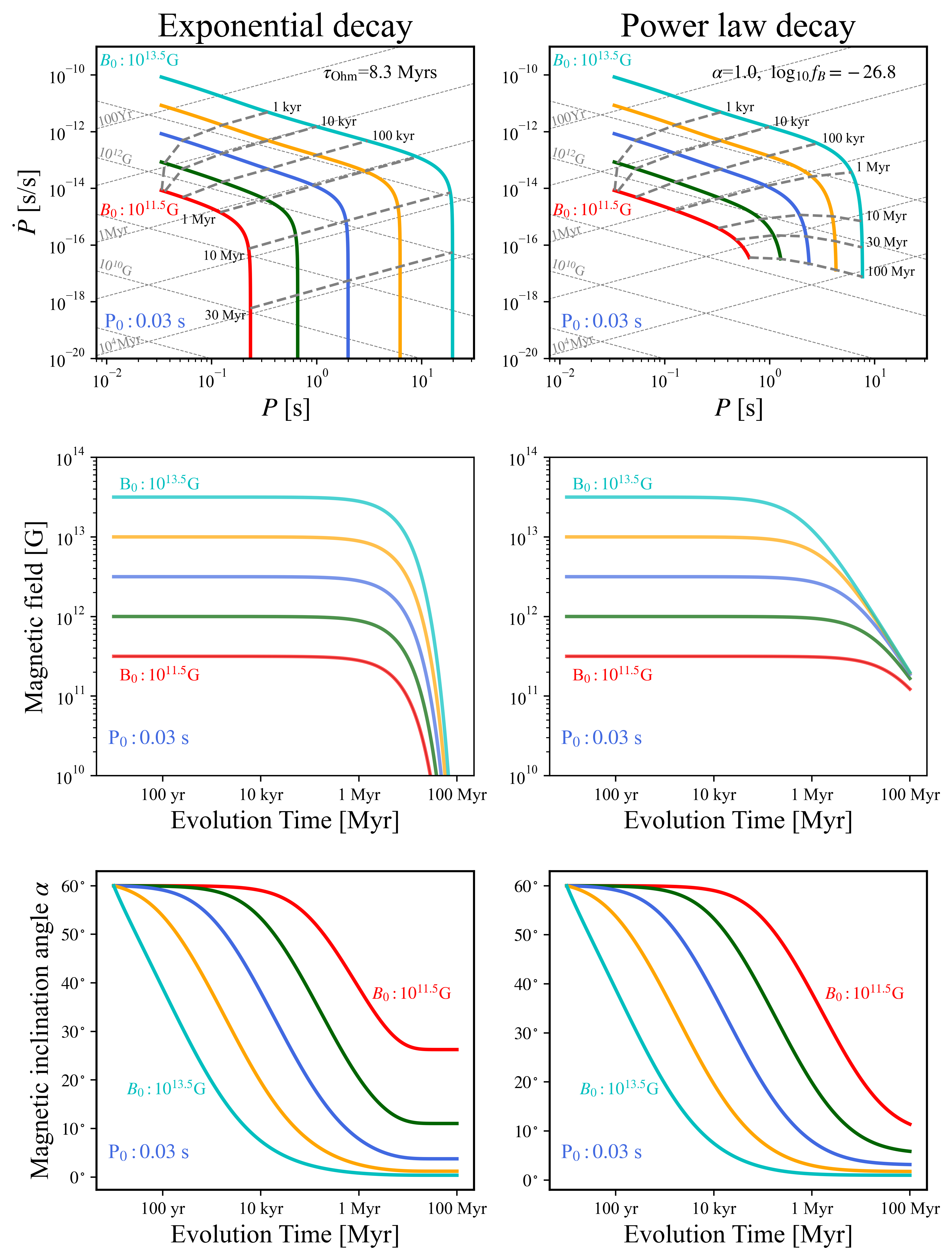}
    \caption{Pulsar evolution tracks in the \ppdot\ diagram and the evolution of the magnetic field and the magnetic inclination angle ($\alpha$) for the best-fit exponential $B$-field decay model. All pulsars start with $P_0=0.03$\,s and initial $\alpha=60\arcdeg$. Different colors lines represent different $B_0$ from $10^{11.5}$\,G to
    $10^{13.5}$\,G, uniform in logarithmic space.
    \label{fig:track}}
\end{figure*}
\subsection{Magneto-rotational evolution}\label{ch:tracks}
To investigate the effect of magnetic field decay on pulsar magneto-rotational evolution, we plot in Figure~\ref{fig:track} the pulsar evolutionary tracks in the \ppdot\ diagram and the time evolution of $B$-field strength and inclination angle. These are calculated using the best-fit exponential decay model with $\tau_{\rm
Ohm}=8.3$\,Myr. We set $P_0=0.03$\,s and initial $\alpha=60\arcdeg$, and try different initial $B$-fields from $B_0=10^{11.5}$\,G to $10^{13.5}$\,G. For illustration purposes, we also plot the power-law decay model with $\alpha_B=1.0$ and $\log f_B=-26.8$.

As shown in the top panel of the figure, the tracks in both scenarios show similar behavior at the beginning. They all follow straight lines of constant $B$-field, since the field decay is not significant at early times. For the exponential decay model, the tracks turn vertically downward when the pulsars are older than $\tau_{\rm Ohm}$, which is due to $B$-field decay, such that
\pdot\ becomes very small ($\dot P\propto B^2$; see Eq.~\ref{eqn:spindown1}) and hence $P$ stays nearly constant.

The power-law decay model shows a similar overall trend but the detailed evolutionary tracks after the turning point depend on $B_0$. The large $B_0$ tracks start to turn downward earlier. This is because the $B$-field decay timescale depends on $1/B_0$ according to Eq.~\ref{eqn:bdecay}, i.e.\ larger
initial $B$-field decays faster. For this reason, all pulsars end up with similar $B$-fields around $10^{11}$\,G at 100\,Myr, much higher than that in the exponential model. This is clearly seen in the $B$-field evolution plot in Figure~\ref{fig:track}. As a consequence, pulsars show a narrow range of $P$ and \pdot\ after evolution, which is not preferable by the observations. We also note that pulsars evolve much slower in this model. 

In addition to the $B$-field strength, the inclination angle $\alpha$ between the magnetic and spin axes is another key factor that determines \pdot. Its evolution is plotted in Fig.~\ref{fig:track}. It is clear that the evolution depends critically on $B_0$. $\alpha$ drops rapidly for large $B_0$ case but it takes a long time to evolve if $B_0$ is low. This is expected as $\dot\alpha\propto -B^2$ (see Eq.~\ref{eqn:spindown2}). At later times, $\alpha$ stops evolving when the $B$-field decays to very small values. In the exponential decay scenario, this happens when the evolution time exceeds $\tau_{\rm Ohm}$. The final value of $\alpha$ therefore remains relatively large for pulsars started with low $B_0$. This effect, on the other hand, is less prominent for the power-law decay model. It can give significantly smaller $\alpha$ for old pulsars, since the $B$-field stays comparatively larger till the end of the evolution.

\begin{figure}[tbh]
    \centering
    \includegraphics[width=0.45\textwidth]{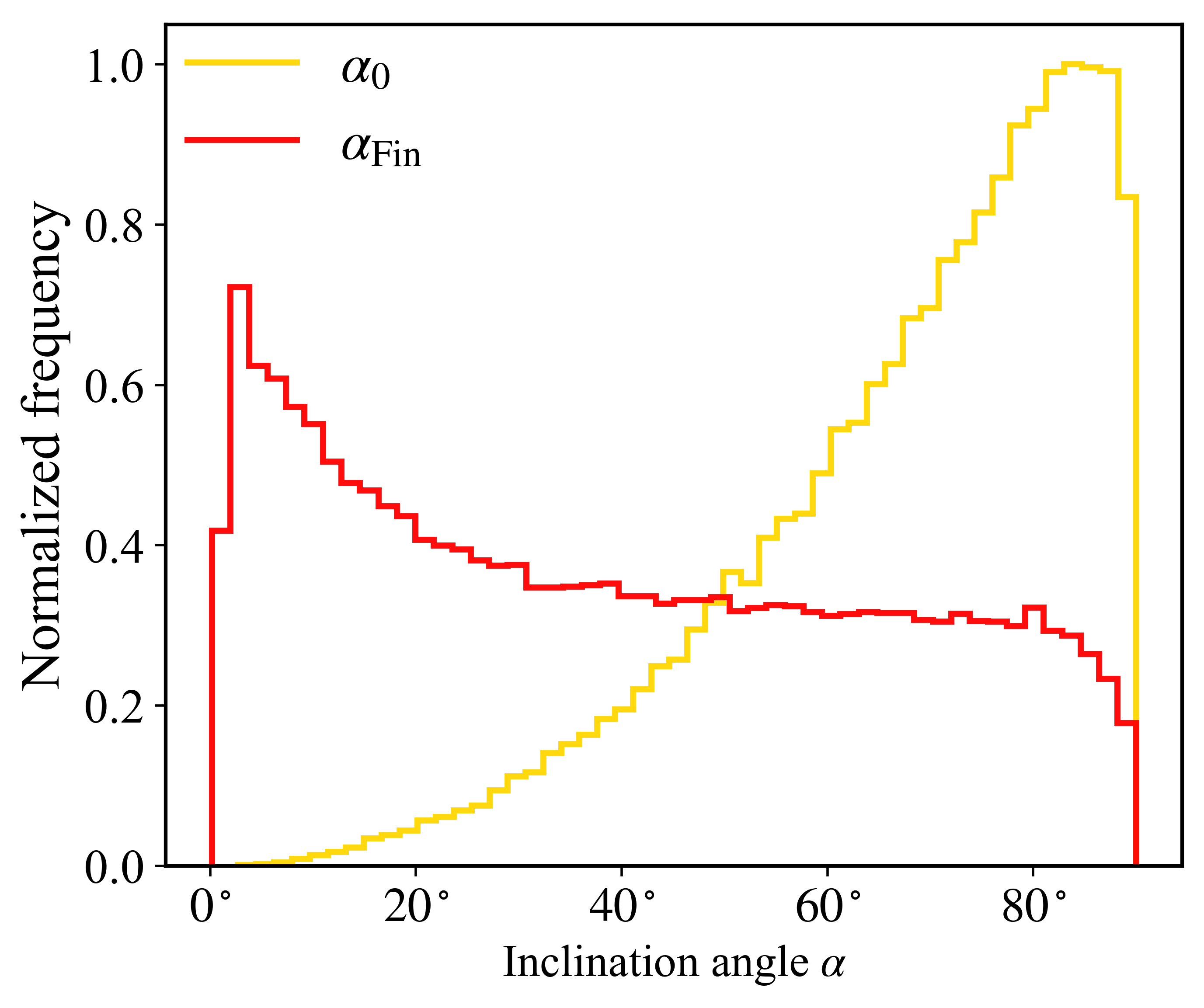}
    \caption{Distributions of the initial and final values (yellow and red lines, respectively)
    of the magnetic inclination angle $\alpha$ for the simulated pulsars detected with the surveys.
    \label{fig:angle}}
\end{figure}
In Figure~\ref{fig:angle} we plot the change of $\alpha$ for the simulated pulsars that are detected by the surveys. The pulsars are assumed to have isotropic $\alpha$ distribution at birth, but those with larger initial $\alpha$ (i.e.\ orthogonal rotators) have a higher chance to be detected. This is mainly due to the geometrical effect as the emission beams of these pulsars are more likely to sweep across the Earth. Moreover, the plot also clearly illustrates the effect of $\alpha$ decay, which makes most of the detected pulsars to become nearly aligned rotators ($\sim$18\% with $\alpha<10\arcdeg$) at the end of the evolution.

\begin{figure}[hbt!]
    \centering
    \includegraphics[width=0.49\textwidth]{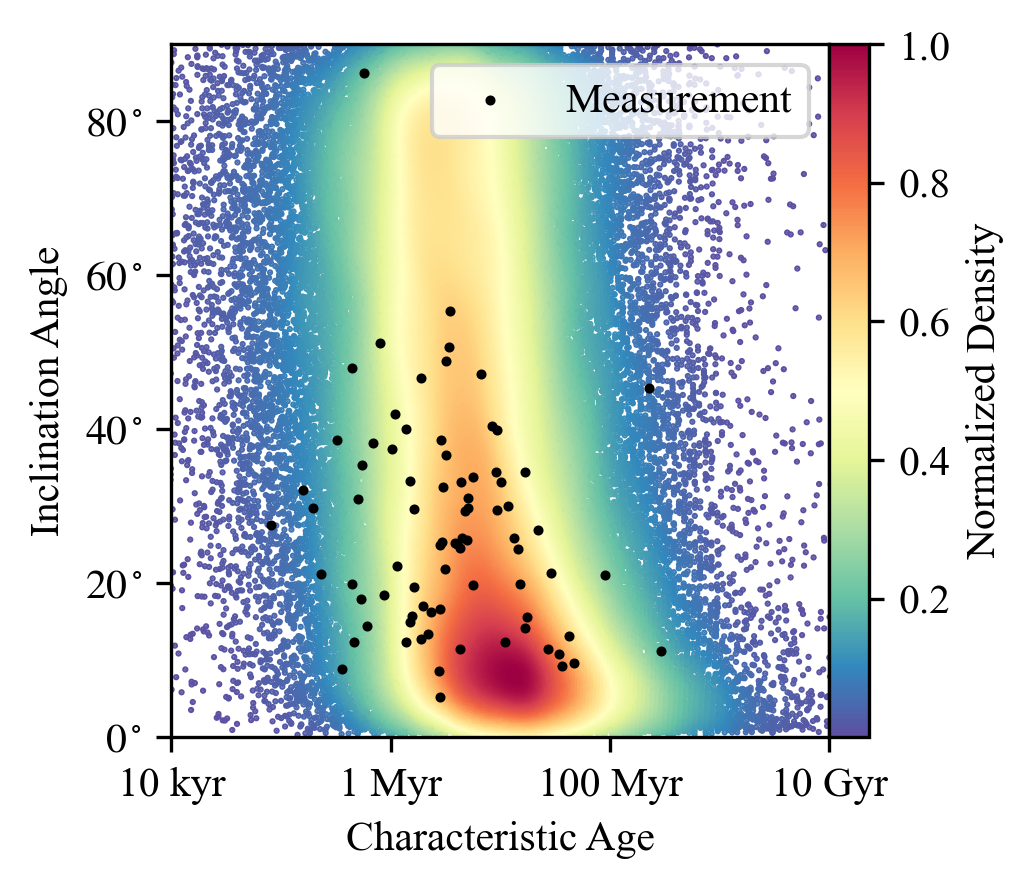}
    \caption{Magnetic inclination angle $\alpha$ versus characteristic age for observed and simulated pulsars. The black dots show 80 measurements of $\alpha$ reported by \citet{malov2011angles}. The underlying density plot is obtained by simulations of 50 times the pulsar sample (i.e. 92950 pulsars) using
    our best-fit population model.
    \label{fig:angle_tau}}
\end{figure}
In Figure~\ref{fig:angle_tau} we plot the observed $\alpha$ versus
characteristic age for 80 pulsars reported by \citet{malov2011angles} and
compare with our best-fit model. The observations are generally consistent
our model prediction, although not many pulsars are found in the most likely
parameter space suggested by the model. We attribute the discrepancy to a small
sample size and also the different beam geometry we adopt (core and cone beam)
than the one (cone beam) assumed by \citet{malov2011angles} to deduce $\alpha$.

\subsection{Gamma-ray pulsar}
After a pulsar is simulated, we calculated its phase-averaged Gamma-ray flux 
according to the equations in Section~\ref{sec:gammaflux}. The result is then
compared with the 14-year \emph{Fermi} all-sky sensitivity
map\footnote{\url{https://fermi.gsfc.nasa.gov/ssc/data/access/lat/3rd_PSR_catalog/}}
\citep{smith2023third} to determine if it is visible. 
Table~\ref{tab:gamma}
lists the number of detectable radio-quiet and radio-loud Gamma-ray pulsars
based on the OG and TPC models, averaged from 100 simulations. We find that the
OG performs better than the TPC. It can produce the number of
radio-loud Gamma-ray pulsars as observed in the Third Fermi-LAT Catalog
\citep{smith2023third}, while the TPC model predicts too many.
Figure~\ref{fig:ppdot} shows a random simulation of Gamma-ray pulsars in the
\ppdot\ diagram using the preferred OG model. They occupy the same parameter
space as the observed pulsars, both are in the upper left of the diagram where
\edot\ is high.

\begin{table}[htb]
\renewcommand{\arraystretch}{1.3}
\centering
\caption{Number of Gamma-ray pulsars in the third Fermi-LAT pulsar catalog
\citep{smith2023third} compared with our simulation results assuming the outer
gap (OG) and the two-pole caustic (TPC) models \citep{watters2009atlas}.}
\begin{tabular}{lcc}
    \hline\hline\vspace*{2pt}
      & Radio-loud & Radio-quiet \vspace*{1pt}\\
    \hline
    3$^{rd}$ Fermi-LAT catalog & 80 (53.3\%) & 70 (46.6\%) \\
    OG model & 81 (59.6\%) & 55 (40.4\%)\\
    TPC model & 128 (69.2\%) & 57 (30.8\%)\\
    \hline
\end{tabular}
\label{tab:gamma}
\end{table}

\section{Discussion \label{s4}}
\subsection{Magnetic field decay model}
In this paper, we present a detailed population study of normal pulsars using realistic physics inputs, including an updated spin-down formula, evolution of magnetic field strength and inclination angle, and sophisticated radio beam geometry. We fit the model to a large sample of
pulsars using a rigorous statistical test and determine the optimal model parameters using MCMC.
We parameterize the $B$-field evolution with the general form $dB/dt\propto B^{(1+\alpha_B)}$ and the best-fit result indicates that $\alpha_B=0$ is preferred. This implies that the exponential decay model can better reproduce the observed pulsar population than the power law. We are able to constrain the decay timescale to be 8.3\,Myr (1$\sigma$ confidence interval of 5.2$-$12.3\,Myr). This is comparable to the mean age of 4.8\,Myr among the detected pulsars in the simulation, implying that $B$-field decay is non-negligible in most pulsars.

It was suggested that Ohmic dissipation in the crust due to phonon scattering is the primary cause of the magnetic field decay in normal pulsars \citep{igoshev2015magnetic} and the Hall effect is only important for $B$-fields stronger than $10^{14}$\,G \citep{vigano2013unifying}, which is 
much higher than that in our simulated pulsars (mean $\log B_0=12.3$ G). In this picture, the $B$-field would show a rapid decay at first driven by the Hall cascade. The Ohmic decay then takes over at later times. This could help explain why the exponential model is preferred as our result shows. The Ohmic decay timescale $\tau_{\rm Ohm}$ depends sensitively on a few physical parameters, including the temperature, impurity factor, and electron fraction \citep{igoshev2021evolution}. Using $T=3.0\times10^{8}$\,K and the typical values for other parameters, we can obtain $\tau_{\rm Ohm}$=8.3\,Myr as our best-fit value.

\subsection{Comparison with previous studies}
Magnetic field decay has been considered in a number of pulsar population synthesis studies. These works adopt different models with various decay timescales ranging from 1\,kyr to 10\,Myr \citep[e.g.,][]{dirson2022galactic,xu2023back,2023arXiv231214848G}.
Only one attempt to constrain the timescale directly from fitting, and a timescale $\tau=4.3\pm0.4$\,Myr was suggested for the exponential decay model \citep{cieslar2020markov}. While this is compatible with our result, we caution that they employ a simple spin-down formula without $\alpha$ evolution and their pulsar sample is much smaller than ours. More importantly, they smoothed the observed pulsar distribution for model comparison, which may not accurately capture the statistic \citep[see][]{2023arXiv231214848G}.

The different decay timescales in previous studies could have an impact on the initial $B$-field required. Our best-fit model prefers slightly lower $B_0$ with $\mu_B=12.3$. This could be attributed to the strong correlation between $\mu_B$ and $\tau_{\rm Ohm}$ as Fig.~\ref{fig:MCMC_Ohm} shows.
Our relatively long time scale suggests a late onset of the $B$-field decay, such that the initial field could be lower. The situation could be more complicated for the power-law model, since the decay timescale would depend on $B_0$ \citep[see][]{dirson2022galactic, xu2023back} instead of fixed as assumed in some works \citep[e.g.,][]{2023arXiv231214848G}.

Overall, all the population models for normal pulsars do not require strong $B_0$ ($\mu_B>13.0$) even with the consideration of field decay. Our best-fit model produces only 0.2\% of pulsars with $B_0>10^{13}$\,G and none above $10^{14}$\,G. This is not sufficient to explain magnetars, which have field strengths of $10^{14}-10^{15}$\,G \citep{kaspi2017magnetars}. This suggests that this class of objects could be a distinct population born with another channel.

The evolution of the magnetic inclination angle was studied in a recent work \citep{dirson2022galactic}, but a slightly different model was adopted for which the decay rate is independent of the $B$-field strength. Similar to our result, it was also found that the effect of $\alpha$ decay is significant, making the majority of pulsars nearly aligned rotators at the end of the evolution.
However, their detected pulsars consist of mainly orthogonal rotators, which is contrary to our result. The discrepancy could be due to different radiation beam models. In particular, our core and cone emission components with Gaussian profiles allow detection even with very small alignment angles.

For the birth spin period distribution, our fit requires similar $\mu_P$ ($\mu_P=-1.40$) as in other studies even using the updated sample (see Fig.~\ref{fig:ppdot}).
However, we note that $\mu_P$ is not very well constrained in the fitting and the large $\sigma_P$ value suggests a broad distribution of $P_0$. Such a result is consistent with the finding that the observed population is insensitive to the $P_0$ distribution and any values in the range of $P_0 < 0.5$\,s could fit well \citep{gonthier2004role, gullon2014population}.

We parameterize the radio luminosity with the general form $L_\nu\propto P^{\alpha_\nu}\dot P^{\beta_\nu}$ as in previous studies. The best-fit result we found $(\alpha_\nu,\beta_\nu)=(-1.76,0.67)$ is close to the typically reported value of $(-1.5,0.5)$ \citep[e.g.,][]{faucher2006birth,bates2014psrpoppy}. Our slightly steeper dependence on $P$ and \pdot\ could be attributed to the $B$-field decay, which results in more aged pulsars in the lower part of the \ppdot\ diagram than in previous works. A larger $\beta_\nu$ is therefore needed to make them non-detectable, in order to match the observations. This also leads to a more negative value of $\alpha_\nu$ due to the strong correlation between $\alpha_\nu$ and $\beta_\nu$ (see Fig.~\ref{fig:MCMC_Ohm}). We note that a very shallow dependence $(-0.39,0.14)$ is recently reported based on the MeerKAT Survey \citep{posselt2023thousand}. This is however not seen in other surveys \citep[e.g.,][]{anumarlapudi2023characterizing} and more observations are needed
to confirm this.

Alternatively, some studies investigate the correlations between $L_\nu$ and \edot\ \cite[e.g.,][]{szary2014radio,wu2020luminosity}. We did not attempt to express $L_\nu$ in terms of \edot\ in our modeling, since this requires a ratio $R_{\alpha\beta}\equiv-\alpha_\nu/\beta_\nu=3$, not supported by our results.
The values of $R_{\alpha\beta}$ reported in the literature always lie between 2 and 3 \citep[e.g.,][]{faucher2006birth,ridley2010isolated,bates2014psrpoppy,johnston2017pulsar,posselt2023thousand}, and the value $R_{\alpha\beta}=2.6$ we obtain is consistent with this range. On the \ppdot\ diagram, this ratio corresponds to the slope of constant $L_\nu$ lines, which can be regarded as
observation-limit lines \citep[see][]{wu2020luminosity}.
It is interesting that the pulsar emission death lines expected from theories also share similar slopes \citep[see][and references therein]{abolmasov2024spin}.
This could imply a deeper physical connection underneath. We however stress that it is highly non-trivial to infer the pulsar intrinsic luminosity from observations, as this requires detailed knowledge of the radio beam geometry, the magnetic inclination angle and the viewing angle. All of these are difficult to obtain and any uncertainties will blur the correlation between $L_\nu$, $P$, and \pdot.

Our study shows that it is possible to reproduce the pulsar \ppdot\ distribution without assuming a death line, same conclusion as \citet{gullon2014population} and \citet{2023arXiv231214848G}. Our luminosity law makes aged pulsars fade away and naturally become undetected. This results in an observation-limit line, which has the same effect as a death line from an observational point of view. On the other hand, we are unable to rule out the death line either. We tried fitting
with the classical death line \citep{bhattacharya1992decay} but there is not much change in the model parameters. This is because in our model most pulsars are already too faint to detect before they reach the death line.
We note that the details of the death line could depend on many physical parameters, including the $B$-field strength, the magnetic inclination $\alpha$ \citep{beskin2022pulsar}, and even the equation of state \citep{zhou2017dependence}. Testing these ideas is beyond the scope of this work.

\subsection{Pulsar birth rate}
Our best-fit population model gives a normal pulsar birth rate of $\sim$0.68 per century in the Galaxy. This value is not affected by the age cutoff of 100\,Myr assumed in the simulation, since pulsars that old cannot be detected.
The birth rate we found is lower than those in previous pulsar
population studies, e.g., 2.8 per century from \citet{faucher2006birth}, 1.4 per
century from \citet{lorimer2006parkes}, and 1.6$-$2 per century from \citet{2023arXiv231214848G}, but still much higher than 0.16 per century from \citet{cieslar2020markov}. Our result indeed better aligns with the recent Galactic core-collapse supernova rate estimate of $1.63\pm0.46$  
\citep{rozwadowska2021rate}. However, we stress that the birth rates inferred from population modeling depend critically on the radio luminosity law, for which many of the details (such as the beam geometry) is not fully understood and
could be subject to large uncertainties as discussed above.
Additionally, including different surveys in the observation sample could give
different birth rates \citep[see][]{2023arXiv231214848G}.
This could be caused by unreliable survey parameters (see Section~\ref{predict} below).


\begin{table}[htb]
\renewcommand{\arraystretch}{1.3}
\centering
\caption{Prediction about the detectable number of future radio surveys. $N_{\rm det}$ and $N_{\rm dis}$ are the number of detectable and newly discovered Galactic normal pulsars by the surveys.}
\begin{tabular}{lcc}
    \hline\hline\vspace*{2pt}
     Survey & $N_{\rm det}$ & $N_{\rm dis}$ \vspace*{1pt}\\
    \hline
    FAST CRAFTS & 1128 & 392 \\
    FAST GPPS & 804 & 418 \\
    SKA & 4646 & 2882\\
    SKA$-$Mid & 2617 & 1073 \\
    MeerKAT TRAPUM & 3604 & 1983\\
    CHIME & 1236 & 498 \\
    \hline
\end{tabular}
\label{tab:predict}
\end{table}

\subsection{Prediction for new surveys\label{predict}}
We employ our best-fit model to make predictions for several upcoming pulsar
surveys: the Commensal Radio Astronomy FAST Survey
\citep[FAST CRAFTS;][]{2011IJMPD..20..989N,li2018fast} with updated
parameters (D. Li; private communication), the FAST Galactic Plane
Pulsar Snapshot survey \citep[FAST GPPS;][]{han2021fast}, 
the Square Kilometre Array survey \citep[SKA and SKA$-$Mid;][]{kramer2015pulsar}
with parameters listed in \citet{cieslar2020markov} and \citet{chakraborty2020understanding},
the MeerKAT TRAPUM survey \citep{ridolfi2021eight}, and the Canadian Hydrogen Intensity Mapping Experiment Pulsar
Survey \citep[CHIME;][]{good2021first}.
The expected number of detections is listed in Table~\ref{tab:predict}. We note that these could be very different from the actual numbers, possibly due to inaccurate survey parameters. 
For instance, the Giant Metrewave Radio Telescope High-Resolution Southern Sky (ghrss) and the Green Bank Northern Celestial Cap (gbncc) surveys were expected to discover 80 new normal pulsars and 130 new MSPs, respectively \citep{bhattacharyya2016gmrt,stovall2014green}, but at the end only 22 and 57, respectively, were found.

\section{Conclusion\label{s5}}
We carry out a population synthesis study for normal radio pulsars employing the most updated
model inputs, including log-normal distributions for $P_0$ and $B_0$ \citep{igoshev2022initial},
the spin-down formula from force-free magnetohydrodynamic simulations with exact dependence on the magnetic inclination angle \citep{philippov2014time}, double Maxwellian distribution for the kick velocity
\citep{verbunt2017observed,igoshev2020observed}, free electron density model by
\cite{yao2017new}, core and cone radio emission model \citep{gonthier2018population}, and latest empirical relation for Gamma-ray luminosity
\citep{kalapotharakos2017fermi,kalapotharakos2019fundamental,kalapotharakos2022fundamental}.
We compared our simulation results with a large pulsar sample that covers all major surveys and performed a fitting utilizing the MCMC approach with 4D KS statistic to constrain the model parameters. Our main findings are:
\begin{enumerate}
\setlength{\itemsep}{-1mm}
\item The observed pulsar population can be well reproduced with the exponential $B$-field decay model. This model also performs substantially better than the power-law decay in terms of goodness of fit. The power-law model can only provide a good fit when $\alpha_B=0$, which reduces into the exponential form. The results suggest that the Ohmic dissipation could play an important part in the $B$-field decay process in neutron stars.

\item Our fitting is able to constrain \textbf{$\mu_B$}\ to be 12.3\ G and the magnetic field decay timescale to be
$8.3^{+3.9}_{-3.0}$\,Myr for the exponential model. This implies that the decay
is non-negligible for most pulsars.

\item
Due to the evolution of the magnetic inclination angle, we expect a large fraction of pulsars to have a good alignment between the magnetic and spin axes.

\item Our best-fit radio luminosity law $L\propto P^{-1.76}\dot P^{0.67}$ can reproduce the pulsar \ppdot\ distribution without the need for a death line. However, we note that the exact values of the indices are model-dependent and the largest uncertainty is from the emission beam geometry.

\item Employing the OG emission model, our population model can well reproduce the number of gamma-ray pulsars observed in the Galaxy.

\end{enumerate}

Our study shows that pulsar population synthesis can provide a powerful tool to probe the physics of neutron stars. In future works, more microphysics could be included, such as the evolution of $B$-field geometry and its dependence on temperature and even the equation of state. 
Also, comparing the simulations with a large observational sample from upcoming surveys will provide better constraints on the model parameters.

\begin{acknowledgments}
We thank Wynn C. G. Ho for the discussion and useful suggestions. We thank the referee for the constructive advice. We thank Di Li for providing us with the FAST\_GPPS survey information. Z.S.\ and C.-Y.N.\ are supported by a GRF grant of the Hong Kong Government under HKU 17303221.
\end{acknowledgments}

\appendix
\section{Comparison with observations}
\label{appendix:a}
\begin{figure}[htb]
    \includegraphics[width=0.49\textwidth]{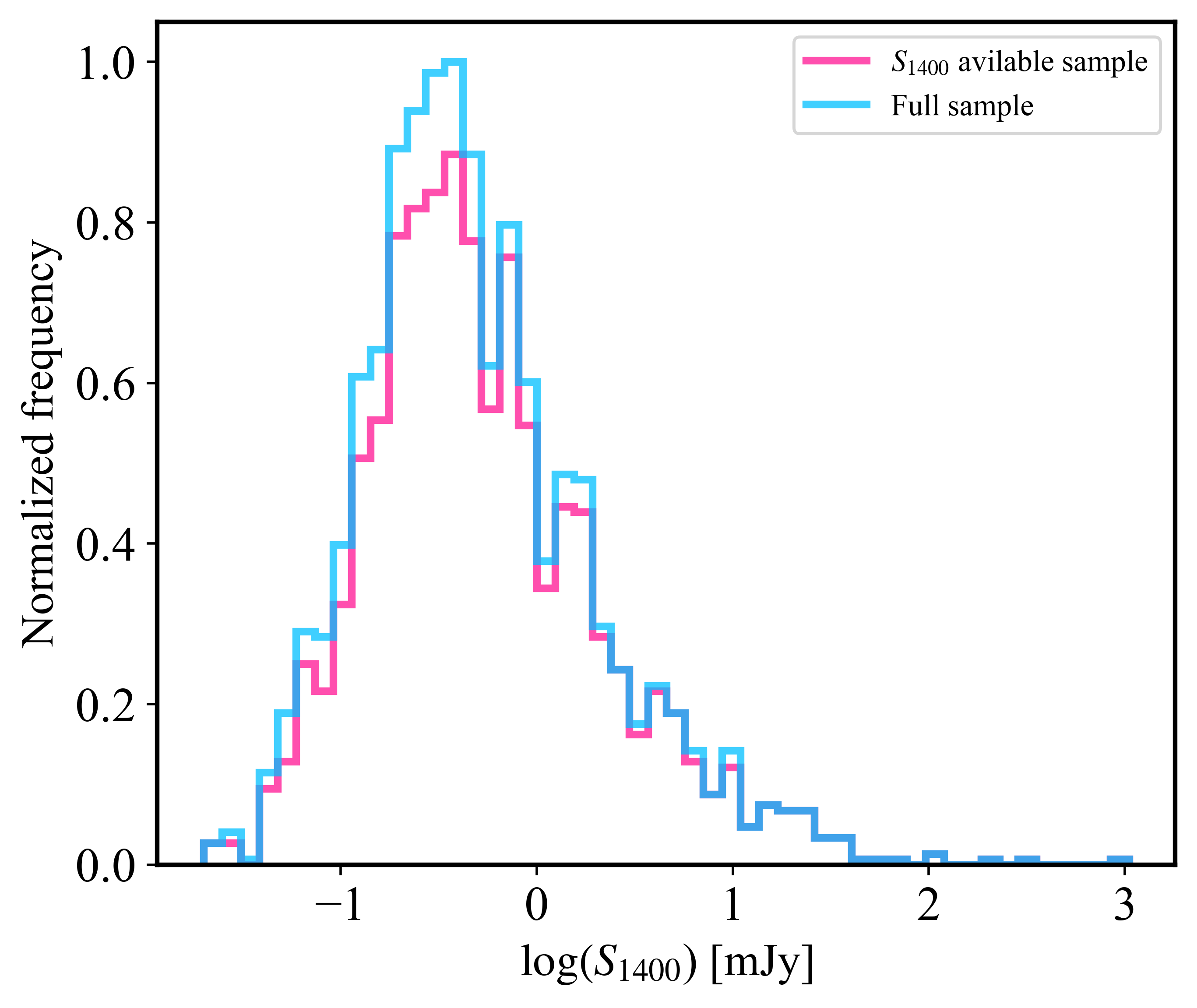}
    \caption{Distribution of radio flux density ($S_{1400}$) for 1624 pulsars with measurements from the ATNF pulsar catalog (red line) compared with the entire sample of 1859 pulsars (blue line).
    For pulsars that have measurements at other frequencies only, we scale their flux densities to 1400\,MHz assuming a spectral index of $-1.60$. 
    \label{fig:sample}}
\end{figure}
\begin{deluxetable}{ll}[!b]
\tablecaption{Radio surveys not used in this work. \label{tab:nosurvey}}
\tablehead{\colhead{Surveys overlapped} & \colhead{Surveys detected fewer}\\[-5pt]
\colhead{with others} & \colhead{than 25 pulsars}\vspace*{2pt}}
\startdata
ar2 (ar3 palfa lotaas) & ghrss (22) \\
ar327 (gbncc) & gbt350 (24) \\
gb1 (pksmb lotaas) & pksngp (13) \\
gb4 (ar4 lotaas pksmb) & pkspa (14)  \\
jb1 (ar3 pksmb lotaas) & pks\_superb (23) \\
jb2 (palfa pksmb) & htru\_eff (13) \\
mol1 (pks70 pkssw pksmb) & pksgc (0) \\
pks1 (pksmb) & tulipp (0) \\
fast\_gpps (ar3 palfa pksmb) & mwa\_smart (0) \\
 & pumps (0) \\
 & meerkat\_trapum (0)
\enddata
\end{deluxetable}
Among the 1859 pulsars in our sample, only 1624 have flux density measurements at 1400\,MHz in the ATNF catalog.
In order to keep those 235 pulsars, we estimate $S_{1400}$ from the measurement at the closest frequency available assuming a power law spectrum with a spectral index of $-1.60$. Figure~\ref{fig:sample} shows a comparison of the flux density distributions with and without these 235 pulsars. We apply the KS test and obtain a statistic of 0.0256 with $p$-value of 0.599, suggesting that using this $S_{1400}$ estimate method has no significant effect on the distribution and hence should not affect our analysis results.

Table~\ref{tab:nosurvey} lists the radio surveys that are not used in this study, either fully overlapped by other surveys (the left column), or detected fewer than 25 normal pulsars according to our selection criteria mentioned in Section~\ref{obssample}. Including all these surveys would only add 38 new pulsars to our sample.

\begin{longrotatetable}
\centering
\begin{deluxetable}{lcccccccccccccc}
\setlength{\tabcolsep}{4pt}
\centering
\tablecaption{Parameters of radio surveys used in this work. The surveys and the corresponding references with survey parameters are as following: Arecibo Survey 1 \citep[ar1;][]{hulse1974high,hulse1975deep}, Arecibo Survey 3 \citep[ar3;][]{fruchter1988millisecond,nice1995search}, Arecibo Survey 4 \citep[ar4;][] {wolszczan1991nearby,thorsett1993search,lorimer2004psr}, Arecibo Multibeam Survey \citep[palfa;][]{cordes2006arecibo}, Princeton-NRAO Survey \citep[gb2;][]{dewey1985search}, Green Bank Short-Period Survey \citep[gb3;][]{manchester1985search}, Green Bank North Celestial Cap Survey \citep[gbncc;][]{stovall2014green}, High Time Resolution Survey-Parkes \citep[htru\_pks;][]{keith2010high} 2$^{\rm nd}$ Molonglo Survey \citep[mol2;][]{manchester1978second}, Parkes High-Latitude Multibeam Pulsar Survey \citep[pkshl;][]{manchester2001parkes,burgay2006parkes}, Parkes Southern Sky Survey \citep[pks70;][]{manchester1996parkes}, Parkes-Swinburne Multibeam Survey \citep[pkssw;][]{edwards2001swinburne}, Parkes Multibeam Pulsar Survey  \citep[pksmb;][]{manchester2001parkes}, LOFAR Tied Array All-sky Survey \citep[lotaas;][]{sanidas2019lofar}. For some old surveys without a clear statement about the survey area in the literature, we estimate the region using their observed pulsar sample listed in the ATNF catalog, and the parameters are labeled with $^\ast$.\label{tab:surveypars}\hspace*{40pt}}
\tablehead{
 \colhead{Parameters} & \colhead{ar1} & \colhead{ar3} & \colhead{ar4} & \colhead{palfa}& \colhead{gb2} & \colhead{gb3} &\colhead{gbncc}& \colhead{htru\_pks}& \colhead{mol2}& \colhead{pkshl}& \colhead{pks70}& \colhead{pkssw}& \colhead{pksmb}& \colhead{lotaas}
\vspace*{5pt}}
\startdata 
    Degradation factor, $\beta$& 1.0 & 1.0 & 1.0 & 1.0 & 1.0 & 1.0 & 1.3 & 1.0  & 1.0 & 3.0 & 1.5 & 1.5 & 1.5 & 1.0 \\
    Antenna gain (K/Jy), $G_0$ & 8--18 & 8--18 & 11 & 8.2 & 1.3 & 1.3 & 2.0 & 0.735 & 5.1 & 0.735 & 0.64 & 0.64 & 0.735 & 1.7\\
    Integration time (s), $t_{\rm obs}$ & 198 & 67.7 & 34 & 268, 134 & 136 & 132 & 120 & 4300, 540, 270 & 100 & 265 & 157.3 & 265 & 2100 & 3600\\
    Sampling time (ms), $\tau_{\rm samp}$ & 5.6 & 0.516 & 0.506 & 0.064 & 16.7 & 2 & 0.08192 & 0.064 & 10  & 0.125 & 0.3 & 0.125 & 0.25 & 0.491\\
    Receiver temperature (K), $T_{\rm rec}$ & 90 & 50--75 & 120 & 24 & 30 & 30  & 46 & 23 & 210 & 21 & 50 & 21 & 21 & 360\\
    Central frequency (MHz), $f$ & 430 & 430 & 429 & 1420 & 390 & 390  & 350 & 1352 & 408 & 1374 & 436 & 1374 & 1374 & 135.25\\
    Bandwidth (MHz), $\Delta F$ & 8 & 10 & 8 & 100, 300 & 16 & 8 & 100 & 340 & 4 & 288 & 32 & 288 & 288 & 31.64\\
    Channel bandwidth (MHz), $\Delta f$& 0.25 & 0.078 & 0.25 & 0.390, 0.293 & 2 & 0.25 & 0.0244 & 0.390 & 0.8 & 3 & 0.125 & 3 & 3 & 0.1953\\
    Minimum RA ($^\circ$) & 0 & 0 & 0 & 0 & 0 & 0 & 0 & 0 & 0 & 0 & 0 & 0 & 0 & 0 \\
    Maximum RA ($^\circ$) & 360 & 360 & 360 & 360 & 360 & 360 & 360 & 360 & 360 & 360 & 360 & 360 & 360 & 360\\
    Minimum Dec ($^\circ$) & $-90$ & $-1$ & $0^\ast$ & $-1$ & $-18$ & $-80$ & $-40$ & $-90$ & $-85$ & $-90$ & $-90$ & $-90$ & $-90$ & $-10$\\
    Maximum Dec ($^\circ$) & +90 & +39 & $36^\ast$ & +38 & +90 & +65 & +90 & +90, +90, +10 &+80 & +90 & 0 & +90 & +90 & +90\\
    Minimum Galactic longitude ($^\circ$) & +35 & +35 & $-180^\ast$ & +32, +170& $-40*$ & $-180$ & $-180$ & $-80$, $-120$, $-180$ & $-180$ & $-140$ & $-180$ & $-100$ & $-130$ & $-180$\\
    Maximum Galactic longitude ($^\circ$) & +60 & +80 & +180* & +77, +210 & $61^\ast$ & +180 & +180 & +30, +30, +180 & +180 & $-100$ & +180 & +50 & +50 & +180\\
    minimum Galactic latitude ($^\circ$) & $-4$ & $-8$ & $-90$ & $-5$ & $-90$ & $-20^\ast$& 0 & $-3.5$, $-15$, $-90$ & $-18$ & $-60$ & $-90$ & $-5$ & $-6$ & $-90$\\
    maximum Galactic latitude ($^\circ$) & +4 & +8 & +90 & +5 & +90 & +20*& +90 & +3.5, +15, +90 & +18 & +60 & +90 & +30 & +6 & +90\\
    Detection SNR threshold, $C$ & 8.5 & 8.0 & 8.0 & 9.0 & 11.0 & 11.0 & 15.0 & 8.0 & 6.3 & 8.0 & 8.0 & 8.0 & 8.0 & 10.0\vspace*{5pt}
    \enddata
\end{deluxetable}
\end{longrotatetable}

\section{Multivariate KS statistic}
\label{Appendix:B}
After each simulation run, we need to compare the distributions of $M$ simulated
pulsars ($M=10N$ in this study) and $N$ observed pulsars ($N=1859$ in our
sample) for $k$ observables $(A, B, C, D, \ldots)$ ($k=4$ in our case).
We evaluate the goodness of fit using the multivariate KS statistic
\citep{fasano1987multid, justel1997multivariate}, which is
calculated as the following:
\begin{enumerate}

\item
\label{pt1}
For each simulated or observed pulsar, call it the $i$-th pulsar ($i=1,\ldots,N$ from observation or $i=1,\ldots,M$ from simulation), which has the properties: ($A_i, B_i, C_i, D_i, \ldots$). We first calculate the cumulative fraction of the observation sample compared with this $i$-th pulsar ($\mathcal{O}_{i}$): 
if $n$ observation pulsars that satisfy the criterion ($A<A_i, B<B_i, C<C_i, D<D_i, \ldots$), then
\begin{equation}
    \mathcal{O}_{i,1}\equiv\frac{n}{N}
\end{equation}
There are totally $2^k$ different values of $\mathcal{O}_i$: $\mathcal{O}_{i,1}$($A<A_i, B<B_i, C<C_i, D<D_i, \ldots$), $\mathcal{O}_{i,2}$($A<A_i, B>B_i, C<C_i, D<D_i, \ldots$), $\ldots$, $\mathcal{O}_{i,2^k}$($A>A_i,\ B>B_i,\ C>C_i,\ D>D_i, \ldots$) accounting for all the different definitions of $\mathcal{O}_i$, and $0\leq \mathcal{O}_i\leq 1$.

\item
Repeat similar step calculating the cumulative fraction of the simulation sample compared with the $i$-th pulsar:
if $m$ simulation pulsars that satisfy the criterion ($A<A_i, B<B_i, C<C_i, D<D_i, \ldots$), then
\begin{equation}
    \mathcal{S}_{i,1}\equiv\frac{m}{M}
\end{equation}
Again there are totally $2^k$ different values of $\mathcal{S}_i$: $\mathcal{S}_{i,1}(A<A_i, B<B_i, C<C_i, D<D_i, \dots$), $\mathcal{S}_{i,2}(A<A_i, B>B_i, C<C_i, D<D_i, \ldots)$, \ldots , $\mathcal{S}_{i,2^k}(A>A_i, B>B_i, C>C_i, D>D_i, \dots)$, and $0\leq\mathcal{S}_i\leq 1$.
\label{itema}

\item
\label{pt3}
For this $i$-th pulsar, calculate the maximum distance between $2^k$ $\mathcal{O}_i$ and corresponding $2^k$ $\mathcal{S}_i$ as $D_i\equiv \max_{(j=1,2,\ldots,2^k)}\left|\mathcal{O}_{i,j}-\mathcal{S}_{i,j} \right|$. If this $i$-th pulsar is from observation, label the distance as $D_{\rm obs,i}$. If this $i$-th pulsar is from simulation, label the distance as $D_{\rm sim,i}$.

\item 
Repeat (\ref{pt1})--(\ref{pt3}) above to obtain the distance for each observed pulsar $D_{\rm obs,i}$ ($i=1, 2,\ldots, N$) and each simulated pulsar $D_{\rm sim,i}$ ($i=1, 2,\ldots, M$). Then calculate the maximum distance among the observed sample and the simulated sample respectively: $D_{\rm obs,max}\equiv\max
(D_{\rm obs,1}, D_{\rm obs,2},\ldots, D_{\rm obs,N})$ and $D_{\rm sim,max}\equiv\max
(D_{\rm sim,1}, D_{\rm sim,2}, \ldots, D_{\rm sim,M})$.

\item The KS statistic value is then $D_{\rm ks}=(D_{\rm obs,max}+D_{\rm sim,max})/2$,
with $0\leq D_{\rm ks}\leq 1$.
\end{enumerate}

The value of $D_{\rm ks}$ indicates how well the simulated pulsar population matches that of the observed sample, in terms of the observables i.e.\ $P$, \pdot, DM, and S$_{1400}$ in this work. The smaller the value the better the match. 

\begin{figure}[htbp]
\includegraphics[width=0.49\textwidth]{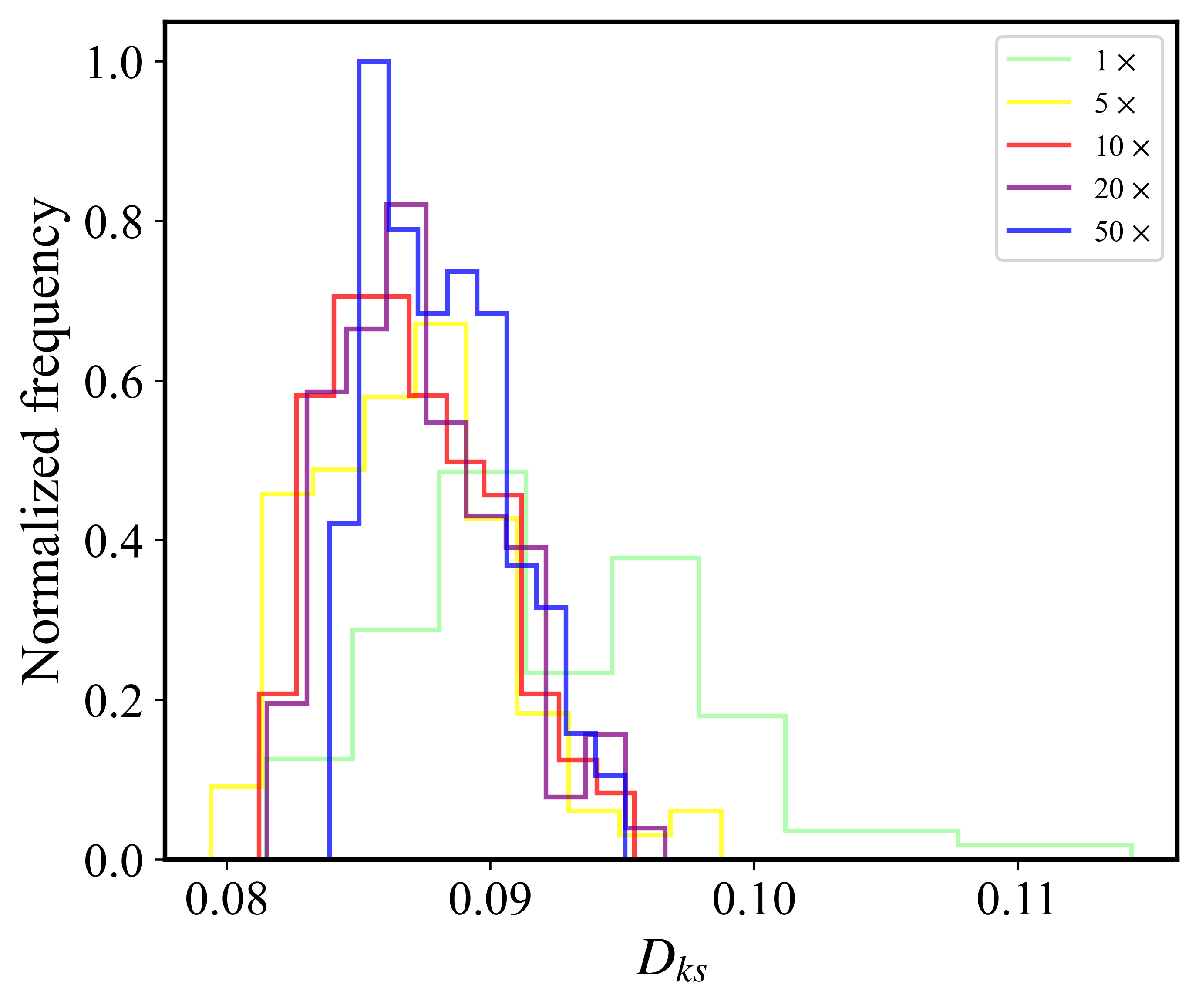}
    \caption{Fluctuation of KS statistic values for 100 simulations, all performed using the best-fit exponential decay model but with a different number of simulated pulsars, from 1 to 50 times as the observation sample. }
    \label{fig:fluc}
\end{figure}

Since our simulation involves random sampling processes, this leads to fluctuation of the KS statistic, i.e.\ slightly different $D_{\rm ks}$ values are obtained every time even with identical model parameters. To investigate this effect, we run simulations with the exponential decay model and the best-fit parameters, and generate 1, 5, 10, 20, 50 times of simulated pulsar as the observation sample.
The whole process is repeated for 100 times to determine the fluctuation in $D_{\rm ks}$.
The result is shown in Figure~A\ref{fig:fluc} and the 1-$\sigma$ (i.e.\ 68\%) range and standard deviations are listed in Table~A\ref{tab:fluc}.
We find that simulating ten times of pulsars as the observation sample (i.e.\ 18590 simulations for 1859 pulsars observed) gives sufficiently small fluctuation of $D_{\rm ks}$ and more simulations provide no significant improvement.
Therefore this is adopted in our simulation.

\begin{table}
\centering
\renewcommand{\arraystretch}{1.2}
\caption{1$\sigma$ range and standard deviation (SD) of KS statistic obtained from 100 simulations, each with 1, 3, 5, 10, 20, 50 times of pulsars as the observation sample. All simulations are done with the exponential model using the best-fit parameters. }
\begin{center}
\begin{tabular}{ccc}
    \hline\hline\vspace*{3pt}
    No.\ of sim.\ pulsars & 1-$\sigma$ range & SD ($\times 10^{-3}$) \vspace*{1pt}\\
    \hline
    $1\times$  & 0.0143 & 5.93 \\
    $3\times$  & 0.0096 & 3.66 \\
    $5\times$  & 0.0086 & 3.62 \\
    $10\times$ & 0.0073 & 3.06 \\
    $20\times$ & 0.0072 & 3.05 \\
    $50\times$ & 0.0058 & 2.57 \\
    \hline
\end{tabular}
\end{center}
\label{tab:fluc}
\end{table}

\clearpage
\bibliography{main}
\end{document}